\documentclass[11pt]{article}
 \newif\ifjv\jvtrue  
\newif\iflncs\lncsfalse
\newif\ifieee\ieeefalse
\newif\ifels\elsfalse

\usepackage[usenames]{color} 
\usepackage{latexsym}
\usepackage{amsfonts}
\usepackage{amssymb}
\usepackage{verbatim}
\usepackage{enumerate}
\usepackage{alltt}  
\usepackage{url}
\usepackage[all]{xy}
\usepackage{array}
\usepackage{xspace}
\ifieee
\usepackage[cmex10]{amsmath}
\else
\usepackage{amsmath}
\fi

\iflncs
\else
\usepackage{amsthm}
\fi

\newcommand{\com}{\newcommand}

\renewcommand{\vec}[1]{\mathbf{#1}}
\com{\ints}{{\mathbb Z}}
\com{\deqarrow}{{\downarrow}%
  {\raisebox{1ex}{\rmfamily\hspace{-1.3ex}\scriptsize=}}}

\newcommand{\trans}[3]{{#2,#3\models{#1}}}
\newcommand{\F}[1]{F^{\mathcal #1}}

\newcommand{\finit}{{f^\iota}}
\newcommand{\setinit}{F_{\textsc{init}}}
\newcommand{\maxi}{\max x^\iota}
\newcommand{\mini}{\min x^\iota}
\newcommand{\ourmcs}{{S(I(\mathcal{A}))}}
\newcommand{\elabmcs}{{\mathcal E}}
\newcommand{\spfl}{{\mbox{\textsc{SFPL}}}}
\newcommand{\abst}{{\mathbb A}}

\newcommand{\MCTS}{\ensuremath{(\mathcal{MC},\ints)\textrm{-CTS}}\xspace}
\newcommand{\CTS}[1]{\ensuremath{(#1)\textrm{-CTS}}\xspace}
\com\clos[1]{{cl(#1)}}
\com\closATf{{\clos{\elabmcs}{\upharpoonright}f}}
\let\eps=\varepsilon
\newcommand\tuple[1]{\langle #1 \rangle}
\newcommand\Aff{\text{\it Aff}}
\newcommand\Gap{\text{\it GC}}

\com{\pgt}[1]{{\tt #1}}
\com\powerset{{\cal P}}
\com{\eqdef}{\stackrel{\mathrm{def}}{=}}
\com{\notvdash}{\not{\!\vdash}}

\iflncs
\spnewtheorem{exmp}{Example}[section]{\bfseries}{\rmfamily}
\spnewtheorem{alg}{Algorithm}[section]{\bfseries}{\rmfamily}
\else
\ifieee
\newtheorem{theorem}{THEOREM}
\else
\newtheorem{theorem}{THEOREM}[section]
\fi
\newtheorem{obs}[theorem]{OBSERVATION}

\newtheorem{lemma}[theorem]{LEMMA}
\newtheorem{claim}[theorem]{CLAIM}
\theoremstyle{definition}
\newtheorem{definition}[theorem]{Definition}

\ifieee
\newtheorem{alg}{Algorithm}
\else
\newtheorem{alg}{Algorithm}[section]
\fi
\theoremstyle{remark}
\ifieee
\newtheorem{exmp}{Example}
\else
\newtheorem{exmp}{Example}[section]
\fi
\fi

\com{\bthm}{\begin{theorem}}
\com{\ethm}{\end{theorem}}
\com{\bdfn}{\begin{definition}}
\com{\edfn}{\end{definition}}
\com{\blem}{\begin{lemma}}
\com{\elem}{\end{lemma}}
\com{\bex}{\bigskip\par\begin{exmp}}
\com{\eex}{\end{exmp}}
\ifieee
\com{\bprf}{\begin{IEEEproof}}
\com{\eprf}{\end{IEEEproof}}
\else
\com{\bprf}{\begin{proof}}
\iflncs
\com{\eprf}{\qed\end{proof}}
\else
\com{\eprf}{\end{proof}}
\fi
\fi

\newcommand{\bi}{\begin{itemize}}
\newcommand{\ei}{\end{itemize}}
\newcommand{\be}{\begin{enumerate}}
\newcommand{\ee}{\end{enumerate}}

\com{\fl}{\noindent}
\com{\hair}{\hspace{3mm}}
\com{\vair}{\vspace{3mm}}

\newenvironment{fig0}[3]
{
\xdef\fighack{\noexpand\caption{#1}\noexpand\label{#3}}
\begin{figure}[#2]
\hspace*{3mm}\begin{minipage}{0.95\columnwidth}}{\end{minipage}%
\fighack%
\end{figure}}

\newenvironment{program}{
\medskip\par 
\begin{minipage}{0.9\textwidth}
\begin{alltt}}{
\end{alltt}
\end{minipage}\par\medskip\par}

\title{Bounded Termination of Monotonicity-Constraint Transition Systems}

\ifieee
\author{\IEEEauthorblockN{Amir M. Ben-Amram and Michael Vainer}
\IEEEauthorblockA{Tel-Aviv Academic College\\
Tel-Aviv, Israel\\
Email: amirben@mta.ac.il}
}
\else
\ifels
\author[aba]{
   Amir M. Ben-Amram%
   \corref{cor1}\fnref{t1}%
}
\cortext[cor1]{Corresponding author}
\address[aba]{School of Computer Science, The Academic College of Tel-Aviv Yaffo,
  PO Box 8401, 61083 Jaffa, Israel.}
\fntext[t1]{Part of this work was done while Amir Ben-Amram was visiting DIKU, the department of Computer 
Science at the University of Copenhagen.}
\ead[aba]{amirben@mta.ac.il}
\author[mv]{Michael Vainer}
\ead[mv]{michaelvainer@gmail.com}
\else
\author{Amir M. Ben-Amram\thanks{School of Computer Science, The Academic College of Tel-Aviv Yaffo,
  PO Box 8401, 61083 Jaffa, Israel.}\ \ and Michael Vainer}
\fi
\fi

\begin{document}

\ifels
\begin{frontmatter}
\else
\maketitle
\fi

\begin{abstract}
Intuitively, if we can prove that a program terminates, we expect some 
conclusion regarding its complexity. But the passage from termination proofs to complexity bounds is not always clear.
In this work we consider Monotonicity Constraint Transition Systems, a program
abstraction where termination is decidable. We show that these programs also have a decidable complexity property: one can determine whether
the length of all transition sequences can be bounded in terms of the initial state. This is the 
\emph{bounded termination} problem. 
Interestingly, if a bound exists, it must be polynomial.
We prove that the bounded termination problem is PSPACE-complete and if a bound
exists then it is polynomial in the initial values.

We also discuss, theoretically, the use of bounds on the abstract program to infer conclusions on a concrete program that has been abstracted.
The conclusion maybe a polynomial time bound, or in other cases polynomial space or exponential time.
We argue that the monotonicity-constraint abstraction promises to be useful for practical complexity analysis of programs.
\end{abstract}

\newcommand\mykeywords{Complexity analysis; Size-change termination; Bound analysis}
\ifieee
\fi
\ifels

\end{frontmatter}
\fi

\let\include=\input

\section{Introduction}
\label{sec:intro}

\paragraph*{On Complexity Analysis of programs}
Automatically inferring complexity properties of computer programs is a
well-established subfield of static analysis (Section~\ref{sec:rw}, \emph{Related Work},
will provide bibliographic references).
 The topic received renewed attention from
static analysis researchers in recent years, sometimes going by the name \emph{cost analysis}, 
\emph{bound analysis} or \emph{growth-rate analysis}.  The overall goal is to develop
algorithms that can process a subject program and answer questions about its computational complexity, namely
its consumption of some resource such as running time, memory usage,
stack usage, etc.  

It is well-known that in the analysis of algorithms, questions about precise running time (in physical units) are
usually abandoned, since studying this measure involves many properties of complex hardware systems as well
as the software platform, which shift the focus from the algorithm itself. In program analysis, one can distinguish
works that concentrate on the real-time dimension (often going by the keyword WCET---worst-case execution time
analysis), and works that concentrate on more robust (and abstract) \emph{program-based}
measures such as number of instructions executed or
just the number of iterations of a loop. Naturally, some works involve both, to varying degrees, however our work
only addresses the program-based analysis.

A typical question that a program analyzer for complexity may be asked to answer is: give an expression of the cost
(say, execution time---which we shall understand as the number of program steps) in terms of (some designated) input values.

Since we are not measuring real time anyway, it seems reasonable, as in algorithm textbooks, to neglect input-independent constants
and use the $O$-notation. This simplifies the problem, but does not change the basic challenge. Even if we only ask for a \emph{complexity class},
for example to separate polynomial-time programs from super-polynomial ones, this problem is still undecidable in every Turing-complete
programming language. This means that there is no hope to solve the problem! How can an algorithm designer overcome such an obstacle?
We list a few alternative approaches (in the context of complexity analysis).

 $\bullet$ 
Focus on specially-designed languages. Such works often grew out
of the research on Implicit Computational Complexity (ICC). In fact, a typical result in this field is
the proof that a complexity class is precisely captured by a particular sub-recursive (Turing incomplete) programming language.
But these languages force the user to program in a particular way, often too unnatural.
Other works show that for suitably restricted languages, the complexity classification is not predetermined but is decidable.
This is an advantage, as it means that the language is less restricted and a more natural programming style should be possible.
\par $\bullet$ 
Give up a complete solution to the problem. This is actually the common approach in the field of \emph{static analysis}, since research
in this field often takes the programming language for given. One then produces analyses that can have ``false negatives" or ``false positives";
in complexity analysis, the most common goal is to provide an \emph{upper bound}, thus the question ``is the program polynomial?'' will
occasionally be answered by a false negative, resulting of an overshot upper bound.
\par $\bullet$ 
A third approach---perhaps a middle road---may be described as \emph{abstract and conquer}. The idea is to first translate a program
from its original language into an \emph{abstract form}, and then analyze the abstract form; a useful abstraction captures important 
aspects of the source program, but it is in the nature of abstraction to lose some precision. One may hope, then, that for abstract
programs one really can \emph{solve} the problem of interest. This may require the development of a good definition of the analysis goals
in the abstract world. This approach can already be seen in different fields of program analysis, including complexity analysis, as we will
mention in more detail below. It has several benefits, in particular, \emph{theoretically}, the abstract program model may be sufficiently
simple to develop a firm theoretical understanding; as problems may be decidable, one can may be able to progress to proving their
computational complexity. \emph{Practically}, the approach suggests a separation of concerns among a front end
and a back end, and promotes modularity in tool construction.

\paragraph*{Termination Analysis}
Termination Analysis is another much-studied topic in program analysis.
Intuitively, a termination proof seems likely to reveal something about the complexity of the subject program,
since if we can explain why a computation progresses towards its end, perhaps we can say how fast it progresses.
It is, therefore, natural to try to extend work on termination proofs to obtain complexity bounds.
In fact, some works on complexity analysis have already exploited techniques from
termination analysis (polynomial interpretation of terms in~\cite{BCMT:01,CichonLescanne:1992}; ranking functions in
\cite{AAGP:sas08,ADFG:2010}). 
In this work too our goal was to examine certain theoretical and algorithmic results from termination analysis
and evolve them to obtain results in complexity analysis. 
Specifically, we study the \emph{monotonicity constraint} abstraction, described below.

\paragraph*{Constraint Transition Systems}
A \emph{Constraint Transition System} (CTS) is an abstract program which is based on viewing
the semantics of the program as an infinite-state transition system which has a finite description.
The components of this description are: first, a  \emph{control flow graph} (CFG), which is a finite directed graph;
we refer to its nodes as flow points. Typically, they represent concrete locations in the source code of the subject program.
Second, a finite set of \emph{variables} associated with every flow point; a state is
specified by $(\ell,x_1,\dots,x_n)$ where $\ell$ 
is a flow point and $x_1,\dots,x_n$ the values of the variables. 
The variables may represent actual program data, abstractions like the size of an object (a list, a tree, a set etc.), in some cases
program constants, and in some cases ``invented'' variables (created by the analysis tool).
Finally, every arc of the CFG, to which we refer as \emph{an abstract transition}, is associated with
a formula that represents a relation on source states and target states (the transition relation).
We refer to this formula as a \emph{constraint}.  A common notation for constraints is to denote the target state
variables by primed identifiers. So, for example, $x>x'$  means that the new value of variable $x$ is smaller than the old one.
Figure~\ref{fig:ex:A} shows a small program and a possible abstraction to constraints (in fact, to monotonicity constraints, as defined below). 
The reader should be able to see that the constraints suffice for deducing that the loop always terminates. Additional examples appear in later sections.

So far, the definition has been very general, and practically any program representation or computational model 
of finite description can be represented in this way. However, certain kinds of CTS are more frequent in program analysis.
To specify a particular kind of CTS, we have to specify the kind of constraints allowed and over what carrier set they work.
In this paper, we employ the notation \CTS{\mathcal C,\mathcal D} for a CTS that applies constraints of type $\mathcal C$
to the domain $\mathcal D$.

\emph{Monotonicity constraints} were introduced to termination analysis as early as 1991~\cite{Sa:91}.
These are constraints that only use order relations $>$ and $\ge$, and their use in termination analysis stems from
the idea of proving termination by identifying a descending sequence---a pattern typical to Logic and Functional programming, where
one often recurses on values such as terms, trees or lists while shrinking them. Hence \emph{size-change termination},
a name given to this approach in~\cite{leejonesbenamram01}. The precise abstraction used in the latter work is this:
\emph{Constraints} are conjunctions of relations of the form $x>y'$ or $x\ge y'$. They are referred to 
as \emph{size-change graphs} (SCG).   Thus, the abstraction employed by
size-change termination
(\`a-la \cite{leejonesbenamram01}) may be expressed as \CTS{SCG,\mathtt{Ord}},
where $\mathtt{Ord}$ stands for ``any well-ordered set."

When one looks at earlier papers using monotonicity constraints (e.g., \cite{Sa:91,LS:97}),
one may notice that their constraint formulae are not restricted to size-change graphs---there was no prohibition of constraints
such as $x<x'$ (an increase, rather than decrease) or $x<y$ (a constraint on source-state variables) or $x'<y'$. We refer to this constraint
domain as $MC$.
 It  also is clear that the intended domain is the
non-negative integers. In 2005, Codish, Lagoon and Stuckey~\cite{Codish-et-al:05} began the extension of size-change termination theory
to monotonicity constraints and the integers.  
To illustrate the need for refining the theory, note that a loop described by the constraint $x<x' \land x<y \land y=y'$,
a common pattern in imperative programs, does not satisfy size-change termination (there are well-ordered sets in which this can be repeated
forever), but terminates over the integers. Note also that when arguing for its termination over the integers, the assumption $x,y\ge 0$ is 
redundant, and in fact in imperative programs the important variables for loop control are often of \emph{integer} type, and can be
(by design or by mistake) negative too.  Note also that the usage of constraints which is not of the ``size-change graph'' type.
This motivated the study of \MCTS in \cite{BA:mcInts}. Two significant results of this study are: 
(1) termination of \MCTS is decidable; it is PSPACE-complete. (2) There is an algorithm for constructing \emph{global ranking functions}
for terminating \MCTS instances.

Other types of CTS have also appeared in termination analysis as well of complexity analysis; more on this below.

\begin{fig0}{CTS abstraction of a simple program (1).}{t}
	       {fig:ex:A}
\setlength{\unitlength}{0.5pt}
\setlength{\extrarowheight}{1ex}
\hspace{1cm}
\begin{tabular}{ll}
{\bf\em Program 1} & {\bf\em CFG and constraints} \\ 
\begin{minipage}{2in}
\ \medskip\par
\fl \verb/while x>z do/ \par
  ~~~ \verb/(x,y) := (y,x-1)/
\end{minipage}
&
$\xymatrix@R=20pt{
& \\
  \circ\ar`d[dr]`[ur]`[u]_{
\setlength{\extrarowheight}{0pt}\scriptstyle
\begin{array}{l}\pgt{x}>\pgt{z} \land\\ \pgt{y}=\pgt{x}' \land\\ \pgt{x}>\pgt{y}'  	\land\\ \pgt{z}=\pgt{z}'\end{array}}%
 `[] []& 
& \\
& 
}$
\end{tabular}
\end{fig0}


\paragraph*{Complexity Analysis of Abstract Programs}
Stated succinctly, a CTS represents a transition relation (relation on the set of states) and the goal of termination analysis is to prove
that this relation is well-founded.  A natural notion of complexity for the abstract program is the (worst-case) number of transitions
starting from an initial state (a state where the program is at its designated point of entry), which we would like to bound in terms of
the variables at that initial state (or a few designated variables).  


Our research on complexity analysis of \MCTS has been inspired by two earlier works on the complexity analysis of programs, which
are both based on a CTS abstraction: the COSTA system of Albert et al.~\cite{AAGP:sas08,AAGP:jar2010}, which targets Java bytecode programs,
 and the WTC analyzer of Alias et al.~\cite{ADFG:2010}, targeting C programs.
For the purpose of this presentation, we
follow~the latter (more on the former in Section~\ref{sec:rw}).
 The \emph{abstraction} used is \CTS{\Aff,\ints} where $\Aff$ denotes a constraint language where a 
constraint is a conjunction of linear (affine) inequalities, for example: $x<1 \land x+y\le z$.  It should be clear that \MCTS is a sub-model
of \CTS{\Aff,\ints}. 
As for \emph{analysis of the abstract program}, the method is to search for a \emph{lexicographic linear ranking function}.
This is a function of the form $\rho_\ell(x_1,\dots,x_n) = \tuple{f_{\ell,1}(\vec x),\dots,f_{\ell,d}(\vec  x)}$ where each $f_{\ell,i}$ is
an affine function on $\ints^n$ whose values in reachable program states $(\ell,\vec x)$ are guaranteed to be non-negative.
Moreover, the value of this function decreases (lexicographically) in every transition. It is easy enough to see that this proves termination;
it may also imply a bound on running time. The bound will be a polynomial of degree  at most $d$
 (the length of the longest tuple used,
also referred to as the dimension). Interestingly, among all functions that satisfy the conditions which
\cite{ADFG:2010} impose on their ranking functions, the algorithm
provably finds one of smallest dimension.

Both of the above works were accompanied by front-ends that abstracted programs, demonstrating the applicability of the approach
to analysis of concrete programs in the respective languages.

The \MCTS abstraction has, previous to our work only been used for proving termination%
\footnote{Concurrently to our work, it was also put to use in complexity analysis by Zuleger et al.~\cite{ZGSV-sas11};
and the bounded termination problem was independently studied in~\cite{Bozzelli:2012}.}%
.
Thus, our first contribution is to define the property of  \emph{bounded termination} in this particular context.
This may seem a trivial step, but introducing this definition was important as it
expressed our realization that \emph{not for every terminating CTS can a complexity bound
be obtained}  (this will be shown precisely in Section~\ref{sec:prelim}).
Hence, the class of bounded-terminating instances is a strict subset of the terminating ones;
which means that the PSPACE-completeness of the set of terminating programs does not mean that the
bounded termination problem also has such complexity---not even that it is decidable.
Our fundamental result is a proof that bounded termination is decidable. Moreover: we prove that it is PSPACE-complete,
the same complexity as for termination; and indeed we re-use some
 techniques from the work on \MCTS termination in both
the upper bound proof and the hardness proof. Unlike~\cite{AAGP:sas08,ADFG:2010,ZGSV-sas11}, we do not use ranking functions.

An interesting consequence of our proofs was the discovery that  bounded termination implies that the bound obtained
 \emph{is always polynomial} (in terms of the  initial values). 
Note that this is an \emph{inherent} property---not an artifact of the analysis algorithm.

\begin{figure}[t]
\setlength{\unitlength}{0.5pt}
\setlength{\extrarowheight}{1ex}
\begin{tabular}{ll}
{\bf\em Program 2} & {\bf\em CFG and constraints} \\ 
\begin{minipage}{2in}
\ \medskip\par
\fl \begin{program}
  i=N; 
  while (i>0) \{
    if (j>0) j--;
    else \{j=N; i--;\}
  \}
\end{program}
\end{minipage}
&
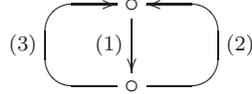
\begin{minipage}{2in}
$\xymatrix@R=20pt{
 & \circ\ar[d]_{(1)}
& \\
  & \circ\ar`r[ur] `[u]_{(2)} [u] \ar`l[ul] `[u]^{(3)} [u]& 
}$ \par
{\small
$\begin{array}{ll}(1) & \pgt{i}>\pgt{0} \land \mbox{Same}(\pgt{N},\pgt{0},\pgt{i},\pgt{j}) \\
(2) & \pgt{j}>\pgt{0} \land \pgt{j}>\pgt{j}' \ \land \mbox{Same}(\pgt{N},\pgt{0},\pgt{i}) \\
(3) & \pgt{j}\le \pgt{0} \land \pgt{j'}=\pgt{N}' \land \pgt{i}>\pgt{i}'  \land \mbox{Same}(\pgt{N},\pgt{0})
\end{array}$
}
\end{minipage}
\end{tabular}
\caption{CTS abstraction of a simple program. The notation
$\mbox{Same}(x,y,\dots)$ is syntactic sugar for indicating abstract variables that are constant in the transition (see Section~\ref{sec:prelim}).}
\label{fig:ex:quadratic1}
\end{figure}

\paragraph*{Monotonicity-constraint systems as a back-end}
Our paper can be viewed as a theoretical study of \MCTS. However, we argue that such constraint transition 
systems are useful as an abstraction of ``real" programs. To support this claim, we have to discuss 
 the manner in which a concrete program is modeled by a {\MCTS}.

In termination analysis, the concrete-abstract connection is always based on the following principle:
\emph{If the concrete program has an infinite execution, the abstract program will have one}. This
is achieved in different ways depending on the nature of the concrete program (e.g., imperative versus
functional).  Complexity analysis complicates this relationship: the above principle clearly does not suffice.
It is therefore necessary to discuss 
what conclusions on the concrete program  may be drawn from bounded termination
of the abstraction.

Section~\ref{sec:programs} is dedicated to this discussion. Our choice is to keep this paper concentrated on
the theory of \MCTS; therefore this discussion is quite informal. The support for our arguments here is not
theorems and proofs, but the practical experience of researchers who, previous to this work, have already
used a CTS abstraction for complexity analysis. We discuss how this abstraction has been done in
\cite{ADFG:2010} and \cite{AAGP:sas08}. The fact that they used a richer constraint language has no
consequence for this discussion.

Briefly,
the simplest case is of an imperative program, without procedure calls.
The CFG of the {\MCTS} is essentially the flow-chart of the program,
and 
the length of the computation is related to time complexity. 

 Next, we consider programs with recursive functions. We argue that
 for such programs, bounded termination most naturally yields a bound on \emph{stack height}.
 Depending on the program's use of ``heap space,'' we may be able to conclude that it runs in 
 polynomial space, or just deduce an exponential time bound.

The fact that our abstraction is coarser than the one used in the cited works \emph{is} relevant to another
concern: the loss of information due to abstraction. 
Section~\ref{sec:realworld} discusses the impact of relaxing the abstraction used in previous works
to \MCTS.  Such relaxation, which may suffice for termination, does not always suffice
for complexity analysis. An example can be seen in Figure~\ref{fig:ex:quadratic1}: for termination, we
could do with a simpler abstraction, eliminating all constraints involving the variable \pgt{N}. But then we would
not obtain a bounded-terminating CTS.

Our thesis is that, despite its relative simplicity, the monotonicity constraint abstraction
stands a good chance of being effective in practice (when used judiciously).
The ultimate test would, of course,
be the construction of an industrial-strength tool; this is far beyond the scope of our work,
but existing related work (see the next section and Section~\ref{sec:realworld}) makes the prospects seem encouraging.

As an additional informal argument to the interest in this abstraction, we include in Section~\ref{sec:examples} a few additional 
examples, collected from previous papers on complexity
analysis, that illustrate different loop behaviours which are still all captured by our model.

A comment in order is that practical cost-analysis tools typically generate explicit bounds, for example, they would 
generate a bound such as $1.5\;n^3-n+2$ or the asymptotic bound $O(n^3)$.  However, the real bound may possibly be $O(n^2)$, since
no tightness is guaranteed. 
 Our algorithms can provide explicit bounds, but they will be definitely over-approximative.
Bounds that have \emph{precise} explicit bounds may be computable, too. We leave this
as a challenge for further research.

\section{Preliminaries}
\label{sec:prelim}

The results in this paper build on previous research on the termination problem of \MCTS.
To make the paper self-contained, we repeat in this section the basic definitions and certain results from 
previous work.

\subsection{Monotonicity Constraint Systems and their semantics}

\bdfn
a {\MCTS} consists of a control-flow graph (CFG),
monotonicity constraints and state invariants, all defined below.
\begin{itemize}
\item
A control-flow graph is a directed graph (allowing parallel arcs)
over the set $F$ of \emph{flow points}. 
Every flow-point $f$ is associated with
a fixed list of \emph{variables}\footnote{Called parameters or arguments in some publications---depending
on the programming paradigm the authors have in mind. Similarly, flow points may be called program points
or locations.}.
The number of variables is called the arity of $f$ and may be denoted by $ar(f)$; the variables themselves
are usually denoted in the text 
by $x_1,\dots,x_{ar(f)}$, though in examples we may use other identifiers, most naturally
the names of variables of the source program. 
\item
A non-empty  set of flow points, $\setinit \subseteq F$, is designated as \emph{initial}.
\item
Every CFG arc $f\to g$  is associated with a
\emph{monotonicity constraint} (MC), being 
a conjunction of  order constraints $x \succ y$ where
 $x,y\in\{x_1,\dots,x_{ar(f)},x_1',\dots,x_{ar(g)}'\}$, and ${\succ}$  is either $>$ or $\ge$; for uniform notation, 
 we also use $\succ^0$ for $\ge$ and $\succ^{-1}$ for $>$.
Note that $<,\le,=$  can be used as syntactic sugar.

 We write $G:f\to g$ to indicate the association of an MC $G$ with its source and target flow-points.
\end{itemize}
\edfn

A calligraphic-style letter (typically $\cal A$, for \emph{abstract program}) is used to denote a {\MCTS}.
$\F{A}$ ($\setinit^{\mathcal A}$) will be its flow-point (initial flow-point) set.  A monotonicity constraint will often be denoted by $G$ because it is
typically represented by a $\emph{graph}$ (as explained below).  
However, when graph-theoretic notions
are applied to $\cal A$ (such as, ``$\cal A$ is strongly connected''),
they concern the underlying CFG.
In the text, a {\MCTS} may be succinctly referred to as ``a system'' when the meaning should be clear.

\paragraph{State Invariants} Our representation of a \MCTS also includes,
for each $f\in F$, an \emph{invariant} $I_f$, which is
a conjunction of order constraints among the variables.
An example is $(x_1 > x_2) \land (x_3 = x_4)$.  It is assumed that these constraints are also included in the
MCs entering or leaving $f$ (note that for an MC entering $f$, the variables will be primed, as they belong to the target state).
This assumption implies that the invariants are only a convenience, a way to indicate that some constraints will hold whenever 
$f$ is visited, irrespective of which of the incoming and outgoing transitions are taken. The reader will see later that our algorithms make significant use
of this information.

\paragraph{Semantics}

Semantically, a {\MCTS} represents a transition relation over a set of (abstract) program states. In a state, every variable
has a specific value.
In this paper, all values are integers (as in~\cite{BA:mcInts} and unlike~\cite[etc.]{BA:mcs,leejonesbenamram01},
which dealt with well-founded sets).

\bdfn[states]
A {\em state\/} of $\cal A$  
is $s=(f,\sigma)$, where
$f\in \F{A}$ and $\sigma:\{1,\dots,n\}\to\ints$ represents
an assignment of values to the variables, where $n=ar(f)$.
 The state is \emph{initial} if $f \in \setinit^{\mathcal A}$.
\edfn

Satisfaction of a predicate $e$ with free variables $x_1,\dots,x_n$ (for example, $x_1>x_2$) by
an assignment $\sigma$ is defined in the natural way, and expressed by
$\sigma\models e$.
If $e$ is a predicate involving the $n+n'$ variables $x_1,\dots,x_n, 
x'_1,\dots,x'_{n'}$, we write $\sigma,\sigma' \models e$ when $e$ is
satisfied by setting the unprimed variables according to
$\sigma$ and the primed ones according to $\sigma'$.

\bdfn[transitions] \label{def:trans}
 A transition is a pair of states, a \emph{source state} $s$ and a \emph{target state} $s'$.
For $G:f\to g\in {\cal A}$, we write $\trans{G}{(f,\sigma)}{(g,\sigma')}$ if
$\sigma,\sigma' \models G$.
 \edfn

Note that we may have unsatisfiable MCs, such as $x_1>x_2 \land {x_2>x_1}$; our algorithms
will identify such MCs and eliminate them from further consideration.

\bdfn[transition system]
The \emph{transition system} associated with ${\cal A}$
is the binary relation $$T_{\cal A} = \{ (s,s') \mid \trans{G}{s}{s'} 
\text{ for some }G\in {\cal A}\}.$$
\edfn

Note that some authors refer to a program representation as a ``transition system.''  We use
this term for a semantic object.  
Our view of a \MCTS is declarative: a set of constraints that describe the transition system ${\cal T}_{\cal A}$.
It is also possible to interpret a \MCTS operationally, as a kind of program.
 Every MC, $G:f\to g$, then represents a step that the program may take when in program location $f$.
 The step consists of non-deterministically choosing values for the primed variables
 such that $G$ is satisfied by the current state plus the chosen new values. The new values are then assigned to the variables, and the
program location changed to $g$.  While we hope that this view may be useful to some readers, our formal development will 
 use the declarative viewpoint.  

\bdfn[run,height]
A {\em run\/} of ${\cal T}_{\cal A}$ is a (finite or infinite) sequence
of states $\tilde s = s_0,s_1,s_2\dots$ such that for all $i>0$ (up to the end of the sequence),
 $(s_{i-1}, s_{i})\in {\cal T}_{\cal A}$.
For a finite run $s_0,s_1,s_2,\dots,s_\ell$ we refer to $\ell$ as its \emph{length}.
The \emph{height} of a state is the length of the longest run beginning at the state.
\edfn

Note that by the definition of ${\cal T}_{\cal A}$, a run is associated
with a sequence of CFG arcs labeled by $G_1,G_2,\dots$ where
$\trans{G_i}{s_{i-1}}{s_i}$. This sequence constitutes a (possibly non-simple) path in the CFG.
As a slight abuse of definition, we may associate the run with $\cal A$ rather than explicitly
mentioning ${\cal T}_{\cal A}$.


\bdfn[termination]
A transition system is \emph{terminating} if it has no infinite run from an initial state.
A {\MCTS} $\cal A$ is \emph{terminating} if ${\cal T}_{\cal A}$ is terminating.
\edfn

This notion of termination was called \emph{rooted termination} in~\cite{BA:mcInts}, which also considered
 \emph{uniform termination}---where reachability from an initial state is not taken into account. In the context
 of work on bounded termination, rooted termination is essential, and therefore the unqualified term will refer,
 in this paper, to rooted termination.

\bdfn[bounded termination]
A transition system satisfies
\emph{bounded termination} if it is terminating and
the height of every initial state is finite.
We say that a {\MCTS} $\cal A$ satisfies bounded termination if ${\cal T}_{\cal A}$ does
(we also say that $\mathcal A$ is \emph{bounded-terminating}).
\edfn

Ben-Amram~\cite{BA:mcInts} proved that {\MCTS} termination is decidable, and, more precisely,
PSPACE-complete. We shall prove the same for bounded termination.  It is important to note that
a terminating program is not necessarily bounded-terminating, as in the next example. Therefore,
the complexity of this decision problem could be different (compare the LOOP programs~\cite{MR:67}, where
complexity analysis is far harder than termination---the latter is trivial while some natural definitions of
the former are undecidable).

\bex \label{ex:Ack}
A classic example of termination analysis is the Ackermann function, here in pure-functional style:
\begin{verbatim}
ack(m,n) = if m<=0 then n+1 else
           if n<=0 then ack(m-1,1) 
           else ack(m-1,ack(m,n-1))
\end{verbatim}
The straightforward abstraction to a \MCTS, has a single-node control-flow graph (the node represents the function
\texttt{ack}), with three self-loops representing the recursive calls
(here in the order of the call sites in the program text):
\begin{align}
&\pgt{m}>\pgt{0} \land \pgt{n}\le\pgt{0} \land \pgt{m}>\pgt{m}' \land \pgt{n'} > \pgt{0}'\land \pgt{0} = \pgt{0}' \\
&\pgt{m}>\pgt{0} \land \pgt{n}>\pgt{0} \land \pgt{m}>\pgt{m}'  \land \pgt{0} = \pgt{0}' \\
&\pgt{m}>\pgt{0} \land \pgt{n}>\pgt{0} \land \pgt{m}=\pgt{m}' \land \pgt{n} > \pgt{n}'\land \pgt{0} = \pgt{0}' 
\end{align}
Note the constraints $ \pgt{0} = \pgt{0}'$; these are included since in our constraint language there is
no notion of constant. Technically, \pgt{0} is a state variable, hence the need for explicitly stating that
it is constant%
\footnote{Constants can also be explicitly added to the constraint language, see \cite{BP2012}.}%
. The need for constraints like that also arises because of the ``frame problem'' (as it is called
in Artificial Intelligence), that is, the need to state explicitly that variables not affected by a transition do not
lose their value. In order to make the writing of these constraints more concise, we use the notation
$\mbox{Same}(x,y,\dots)$ for $x=x' \land y=y' \land \dots$ (as in Figure~\ref{fig:ex:quadratic1}).

Returning to the Ackermann example, it is easy to verify that
this constraint transition system terminates; in fact, it has a lexicographic 
ranking function $\langle \pgt{m},\pgt{n}\rangle$.
 However, it is not bounded-terminating. Indeed, for any (arbitrarily large) number $N$, it
 has a transition sequence of length $N+1$ from the initial state $(2,1)$:
 $$
 (2, 1) \mapsto (1, N) \mapsto (1, N- 1) \mapsto \dots \mapsto (1,0)
 $$
 The concrete program is, of course, bounded-terminating, because it is deterministic. Thus the length of the run is a function
 of the initial state. This information is lost because the abstraction is non-deterministic, 
and super-approximates the semantics of the concrete program.
 To be more precise, it is the fact that we have \emph{unbounded non-determinism}
 that causes the problem; if the abstraction had been non-deterministic, but finitely branching, by K\"onig's lemma it would still
 be bounded-terminating. 
 
 Finally, we may remark that in this example the program actually computes with the
 non-negative integers and the basic form of size-change graphs (as in~\cite{leejonesbenamram01})
 suffices for its analysis. In this paper we focus on the more general
 \MCTS{} abstraction, but if simple SCGs suffice, some computations in our algorithms become
simpler.
\eex

\subsection{MC graphs and multipaths}

It is convenient for reasoning, and practical for algorithms, to represent 
MCs as directed graphs. These graphs have nodes $x_1,\dots,x_n, 
x'_1,\dots,x'_{n'}$ for the appropriate arities $n,n'$
and represent each relation $x\succ y$ by an arc; an arc representing a strict inequality is called a \emph{strict arc}.
A path in the graph is called strict if it includes at least one strict arc.

Standard graph algorithms can be used to perform operations such as path-finding and ensuring that
the representation is  \emph{transitively closed}, which means that if $x$ can reach $y$ via a (strict) path,
there is a (strict) arc $x\to y$. This computation 
is a standard weighted-reachability closure
in the graph. In the process, we also identify (and remove) unsatisfiable MCs. Clearly, an MC is
unsatisfiable if and only if there is a strict cycle.

\bex\label{ex:running}
Figure~\ref{fig:exintro} shows MCs extracted from the  program below. The flow-points are
\pgt{w} (entry to the \pgt{while} command) and \pgt{i} (entry to the \pgt{if} statement). 

\begin{program}
while (m<n)
  if (m>0) n := n-1
  else m := m+1
\end{program}
\eex
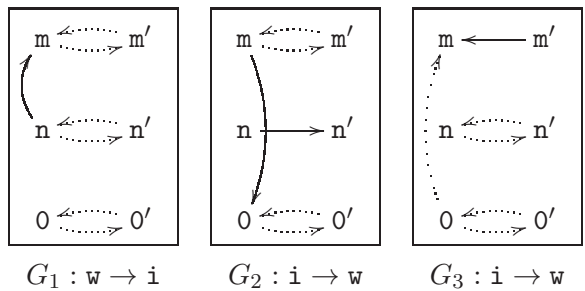
\begin{fig0}{MCs as graphs.
The left-hand side is the source. Broken arcs are non-strict, solid arcs are strict.}{t}
{fig:exintro}
\setlength{\extrarowheight}{1ex}
\begin{centering}
\begin{tabular}{@{\extracolsep{0pt}}ccc}
\fbox{\ $\xymatrix@R=20pt{
  \texttt{m}\ar@/_4pt/@{.>}[r]& \texttt{m}'\ar@/_4pt/@{.>}[l]  \\
  \texttt{n}\ar@/_4pt/@{.>}[r]\ar@/^8pt/@{->}[u] & \texttt{n}' \ar@/_4pt/@{.>}[l]\\
  \texttt{0}\ar@/_4pt/@{.>}[r] & \texttt{0}'\ar@/_4pt/@{.>}[l]
}$\ } &
\fbox{\ $\xymatrix@R=20pt{
  \texttt{m}\ar@/_4pt/@{.>}[r]\ar@/^8pt/@{->}[dd]& \texttt{m}'\ar@/_4pt/@{.>}[l]  \\
  \texttt{n}\ar@{->}[r] & \texttt{n}' \\
  \texttt{0}\ar@/_4pt/@{.>}[r]    & \texttt{0}'\ar@/_4pt/@{.>}[l]
}$\ } &
\fbox{\ $\xymatrix@R=20pt{
  \texttt{m}& \texttt{m}' \ar@{->}[l]\\
  \texttt{n} \ar@/_4pt/@{.>}[r]& \texttt{n}'\ar@/_4pt/@{.>}[l] \\
  \texttt{0}\ar@/_4pt/@{.>}[r]\ar@/^8pt/@{.>}[uu] & \texttt{0}'\ar@/_4pt/@{.>}[l]
}$\ } \\
$G_1: \pgt{w}\to\pgt{i}$ & $G_2:\pgt{i}\to\pgt{w}$ & $G_3:\pgt{i}\to\pgt{w}$ \\
\end{tabular}\par
\end{centering}
\end{fig0}

Some publications use the term MC graph (MCG); we, however,
identify an MC with its graph representation. This should not cause any problems.
We also use set notation, such as $(x > y) \in G$.
We employ the same notations with respect to state invariants, e.g.,  $(x > y) \in I_f$.
Note that the above example does not have any state invariants (in the given abstraction),
because the transitions that enter and exit each point do not agree on any relation among the 
state variables.

\noindent\emph{Notation.} Whenever graphs are considered, the notation $u\leadsto v$ means that there is a path from $u$ to $v$.
 The notation $p:u\leadsto v$ names the path.

\bdfn[multipath]
Let $\cal A$ be a {\MCTS}. Let  $f_0,f_1,\dots \in \F{A}$ be a (finite or infinite) list of flow-points
connected by MCs  ${G_t: f_{t-1}\to f_t}$ (clearly, this constitutes a path in the CFG).
The {\em multipath} $M$ that corresponds to this path
is a (finite or infinite) graph with nodes $x[t,i]$, where $t$ ranges from 0
up to the length of the path (which we also refer to as the length of $M$),
and $1\le i\le ar(f_t)$. Its arc set is the union of the following sets:
for all $t\ge 1$, $M$ includes the arcs of
$G_t$, with source variable $x_i$ renamed to $x[t-1,i]$ and target variable
$x_j'$ renamed to $x[t,j]$. 
\edfn

The multipath may be written concisely as $G_1G_2\dots$;
for example, Figure~\ref{fig-multipath} illustrates a multipath $G_2G_1G_2$, based on the MCs from Figure~\ref{fig:exintro}.
The term \emph{multipath} (originating in~\cite{leejonesbenamram01}) hints at the multiple paths that may exist in the graph representation
of $M$ (the importance of these paths is further discussed below).
We use the expression $\cal A$-multipath when it is necessary to name the CTS that $M$ is formed from.

 If $M_1, M_2$ are finite multipaths, and $M_1$ corresponds to a CFG path that ends at the flow-point where $M_2$
begins, we denote by $M_1M_2$ the result of concatenating them
in the obvious way.  The notation $M:f\leadsto g$ indicates the initial and final flow-points of $M$.

\begin{fig0}{A multipath.}{t}{fig-multipath}
\SelectTips{cm}{}
$$\xymatrix@R=20pt@C=30pt{
  x[0,1]\ar@<2ex>@/^10pt/@{->}[dd]\ar@/^5pt/@{.>}[r] & x[1,1]  \ar@/^5pt/@{.>}[l] \ar@/^5pt/@{.>}[r] & x[2,1]\ar@<2ex>@/^10pt/@{->}[dd]\ar@/^5pt/@{.>}[r]\ar@/^5pt/@{.>}[l] & x[3,1] \ar@/^5pt/@{.>}[l]\\
  x[0,2] \ar@{->}[r]  & x[1,2] \ar@/_5pt/@<-2ex>@{->}[u]\ar@/^5pt/@{.>}[r]  & x[2,2]\ar@/^5pt/@{.>}[l]\ar@{->}[r]
  & x[3,2] \\
  x[0,3]\ar@/_4pt/@{.>}[r] & x[1,3]\ar@/_4pt/@{.>}[l]\ar@/_4pt/@{.>}[r]
  & x[2,3]\ar@/_4pt/@{.>}[l] \ar@/_4pt/@{.>}[r]& x[3,3]\ar@/_4pt/@{.>}[l]
}$$
\end{fig0}

Clearly, a multipath can be interpreted as a conjunction of constraints
on a set of variables associated with its nodes. 
We consider assignments $\sigma$ to these variables, where the value
assigned to $x[t,i]$ is denoted $\sigma[t,i]$.

A multipath may be seen  as an execution trace of the abstract program, whereas
a satisfying assignment constitutes a (concrete) run of ${\cal T}_{\cal A}$. Conversely:
every run of ${\cal T}_{\cal A}$ constitutes a satisfying assignment to the corresponding multipath.
Multipaths that start at an initial flow-point are
called \emph{rooted}.  Termination can thus be expressed as non-existence of
satisfiable, rooted infinite multipaths.  

As for single MC graphs, we have

\begin{obs}
A finite multipath is satisfiable if and only if it does not contain a strict cycle.
\end{obs}

We next consider down-paths and up-paths. The definition of a down-path is just the standard definition
of a graph path, but it is renamed in order to accommodate the notion of an up-path.

\bdfn 
A \emph{down-path} in a graph is a sequence $(v_0,e_1,v_1,e_2,v_2,\dots)$ where
for all $i$, $e_i$ is an arc from $v_{i-1}$ to $v_i$ (in the absence of
parallel arcs, it suffices to list the nodes).
An \emph{up-path} is a sequence
$(v_0,e_1,v_1,e_2,v_2,\dots)$ where
for all $i$, $e_i$ is an arc from $v_{i}$ to $v_{i-1}$.

The term path may be used generically to mean either a down-path or an up-path (such usage should be
clarified by context).
\edfn

Semantically, in an MC or a multipath, a down-path represents a descending chain of values,
whereas an up-path represents an ascending chain. Note also that an up-path listed backwards is a down-path.

\bdfn 
Let $M = G_1G_2\dots$ be a multipath.
A \emph{down-thread} in $M$ is a down-path that only includes arcs 
of the form ($x[t,i] \to x[t+1,j]$).

An \emph{up-thread} in $M$ is an up-path that only includes arcs of the form
($x[t,i] \gets x[t+1,j]$).

A \emph{thread} is either.
\edfn

For example: in Figure~\ref{fig-multipath}, one down-thread is $x[0,2] \to x[1,2] \to x[3,2]$. This is in fact a strict down-thread,
since it includes two strict arcs. One up-thread is $x[0,1] \gets x[1,1] \gets x[2,1]$.
There are many paths that are not threads, e.g., those that include the arc
$x[0,1]\to x[0,3]$, or those that include cycles.

\bdfn[cyclic] 
We say that a transition, a CFG path, or a multipath,
is \emph{cyclic} if its source and target flow-points are equal.
\edfn

The next lemma and the following definitions are all from~\cite{BA:mcInts}.

\blem \label{lem:finiteM}
If a strongly connected {\MCTS} satisfies SCT, every finite multipath
includes a strict, complete thread.
\elem

\bdfn[composition] \label{def-composition}
The {\em composition} of MC $G_1:
f\to g$ with $G_2: g\to h$, written $G_1;G_2$, is a MC with source
{$f$} and target {$h$}, which includes 
all the constraints among $s,s'$ implied by
$\exists s'' : \trans{G_1}{s}{s''}\land \trans{G_2}{s''}{s'}$.
\edfn


\bdfn[collapse] \label{def:collapse}
For a finite multipath $M = G_1\dots G_\ell$,
Let $\overline M = {G_1;\cdots;G_\ell}$. This is called the
\emph{collapse} of $M$.  
\edfn

\bdfn[reachability]
A flow-point $f\in\F{A}$ is \emph{reachable} if there is a satisfiable finite multipath $M:f_0\leadsto f$
such that $f_0$ is initial.
\edfn

\bdfn \label{def:closure}
Given a {\MCTS} $\cal A$,  its closure set $\clos{\cal A}$ is
the set of collapsed multipaths, $\overline M$, where $M$ ranges over satisfiable finite $\cal A$-multipaths that start at a reachable flow-point.
\edfn

\subsection{Stability}

\bdfn[stability]
A {\MCTS}  $\cal A$  is \emph{stable} if 
(1) all MCs in $\cal A$ are satisfiable;
(2) in the CFG of $\cal A$, all flow-points are reachable from an initial flow-point;
(3) to every $f\in\F{A}$ is associated an invariant $I_f$ such that for all 
$G: f \to g$ in $\cal A$,  $(x_i \succ x_j)\in G \;\iff\; ( x_i \succ x_j) \in I_f$;
similarly,  $(x'_i \succ x'_j)\in G\;\iff\; ( x_i \succ x_j) \in I_f$.
\edfn


\blem \label{lem:stable=satisfiable} {\upshape{\cite{BA:mcs}}} Suppose that {\MCTS}  $\cal A$  is stable. Then
every finite multipath is satisfiable.
\elem

Note that in stable systems the flow-point invariants play an essential role since they are supposed to
contain all the information that can be deduced from the adjacent MCs. The process of 
\emph{stabilizing} a {\MCTS} involves splitting flow-points in the CFG whose original invariants
were not precise enough. 
Algorithms for stabilization are described in~\cite{BA:mcs}. 
Such an algorithm transforms a {\MCTS} $\cal A$ into an equivalent stable system, which
we denote by $S(\mathcal{A})$  (``equivalent'' means that they have the same runs, up to renaming of
flow-points or possibly variables).  We say that $S(\mathcal{A})$ is a \emph{refinement} of $\mathcal{A}$, since
it explicitly separates states that in $\mathcal{A}$ are not explicitly separated.

\begin{fig0}{A stabilized CFG.}{t}{fig:stableCFG}
\entrymodifiers={++[F-]}
\SelectTips{cm}{}
\def\objectstyle{\pgt}
$$\xymatrix{ 
w:\  0<m<n \ar@/_4pt/@{->}[r]_{G_1} & i:\  0<m<n \ar@/_4pt/@{->}[l]_{G_2}\ar@{->}[d]_{G_2} \\
i:\  m<n,\ m\le 0 \ar@{->}[r]^{G_3}\ar@{->}[u]_{G_3}\ar@{->}[rd]^{G_3}\ar@/_4pt/@{->}[d]_{G_3} & 
w:\  {m\ge n,\ m>0}  \\
w:\  m<n,\ m\le 0 \ar@/_4pt/@{->}[u]_{G_1} & w:\  n\le m\le 0
}$$
\end{fig0}

Figure~\ref{fig:stableCFG} shows how the CFG of Example~\ref{ex:running} (which originally had two nodes)
is transformed by stabilization. The \pgt{w} node has been split in four and the \pgt{i} node in two.
There are  also several CFG arcs that represent the same original transition, for example $G_1$
appears twice. The MCs annotating
these arcs will not be identical to $G_1$, since the source and target invariants are merged into each MC.
Note also that there are now several initial flow-points (namely all the nodes labeled (\pgt{w})).

In the worst case, such a transformation can multiply the size of the system
by a factor exponential in the number of variables $n$ (bounded by the \emph{Ordered Bell Number} $B_n$ which is between $n!$
and $2^{n-1}n!$~\cite{BA:mcs}).

\section{The Bounded Termination Problem}
\label{sec:mainThm}

This section gives our first theoretical result: decidability and complexity of the bounded termination problem,
and the corollary that height bounds are polynomial.

\subsection{Discovering Bounded Variables}

To establish bounds on transition-sequence length we need bounds on the values of variables
throughout the execution, in terms of the initial values. So, we are looking for invariants of the
kind $x_i \le x_j^\iota$ where $x_j^\iota$ is the \emph{initial} value of $x_j$.  The inequality relates values at two
different points in execution, not a property of a state, which can be captured by a state invariant.
This apparent difficulty is easily solved by \emph{instrumenting} the program.
Specifically, we make a copy of the initial variables. The copies are never modified but carried over to every
subsequent state and turn the relationship of current values to initial values into a property of states.
In this paper we will, for simplicity, create only two such variables: $x_{max}$ to represent the maximum
among initial values, and $x_{min}$ to represent the minimum. This will allow us to determine whether 
a subsequently-computed value is upper-bounded by at least one initial value (which is the same as
being bounded by $x_{max}$) or lower-bounded by at least one initial value (same as lower-bounded by $x_{min}$).
Note that this instrumentation is part of the algorithm whose input is the constraint transition system; 
we do not deal with concrete programs.
We find it more legible to avoid using  numeric indices for these variables, though technically
they will just be $x_{n+1}$ and $x_{n+2}$ where $n$ is the original arity.

\bdfn
For a given {\MCTS} $\cal A$,
the instrumented version $I(\mathcal{A})$ is obtained by the following steps. \\
(1) Add two new variables $x_{max}$, $x_{min}$ to every flow-point. \\
(2) Add a new initial point $f_0$ with an invariant $I_{f_0}$ that expresses
the intended relationship of $x_{max}$ and $x_{min}$ to the initial value of 
$x_j$ for $1\le j \le ar(f_0)$, namely $x_{max} \geq x_j$ and $x_{min}\le x_j$. \\
(3) Add a transition from $f_0$ to each of the original initial points, with constraints $x_i=x_i'$, for all $i$
(in addition to constraints propagated from $I_{f_0}$ or from the target point).
(4) Add constraints $x_{max} = x_{max}'$  and $x_{min}=x_{min}'$ to all transitions.
\edfn

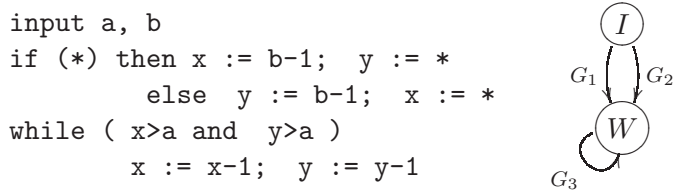
\begin{fig0}{A program and its CFG (with a flow-points for each of the If and
While commands).}{t}{fig:bounding}

\begin{minipage}{7cm}
\begin{verbatim}
input a, b
if (*) then x := b-1;  y := *
         else  y := b-1;  x := * 
while ( x>a and  y>a )
        x := x-1;  y := y-1
\end{verbatim}
\end{minipage}
\parbox{5cm}{
$\xymatrix@R=20pt{
  *++[o][F]{I}\ar@/_/_{G_1}[d]\ar@/^/^{G_2}[d] \\
  *++[o][F]{W}\ar@(l,d)[]_{G_3} \\ 
}$
}
\end{fig0}

As an example, consider the program in Figure~\ref{fig:bounding}, shown together with
its control-flow graph  The notation \pgt{x:=*} represents the assignment of a value unrelated to the program inputs
(such as user input, data from a database etc.), or a value that is not \emph{known} to be so related (e.g., the result
of a function call, or even an arithmetic expression computing a complex function of the current variables).
The invariants and transition constraints for this program are as follows (obtained by manual translation from the
program text):

$
\begin{array}{l@{\ :\ }l}
I_I & \text{true} \\
G_1 & \pgt{x}'<\pgt{b} \land \pgt{b}'=\pgt{b} \land \pgt{a}'=\pgt{a} \\
G_2 & \pgt{y}'<\pgt{b} \land \pgt{b}'=\pgt{b} \land \pgt{a}'=\pgt{a} \\
I_W& \pgt{x}>\pgt{a} \land \pgt{y}>\pgt{a} \land \mbox{Same}(\pgt{x},\pgt{y},\pgt{a},\pgt{b}) \\
G_3 & \pgt{x}'<\pgt{x} \land \pgt{y}'<\pgt{y} \land \pgt{b}'=\pgt{b} \land \pgt{a}'=\pgt{a} 
\end{array}
$

\vspace{\belowdisplayskip}
\noindent
And the instrumented system (with a new flow-point $f_0$ connected to $I$ by a new transition $G_0$):

$
\begin{array}{l@{\ :\ }ll}
I_{f_0} & x_{max} \ge \pgt{x} \land  x_{max} \ge \pgt{y} \land x_{max} \ge \pgt{a} \land x_{max} \ge \pgt{b} & \land\\
\multicolumn{2}{r}{ x_{min} \le \pgt{x} \land x_{min} \le \pgt{y} \land x_{min} \le \pgt{a} \land x_{min} \le \pgt{b}}\\
G_0 & I_{f_0} \land \mbox{Same}(\pgt{x},\pgt{y},\pgt{a},\pgt{b},x_{min},x_{max}) \\
I_I & \text{true} \\
G_1 & \pgt{x}'<\pgt{b} \land \mbox{Same}(\pgt{a},\pgt{b},x_{min},x_{max}) \\
G_2 & \pgt{y}'<\pgt{b} \land \mbox{Same}(\pgt{a},\pgt{b},x_{min},x_{max}) \\
I_W& \pgt{x}>\pgt{a} \land \pgt{y}>\pgt{a} \land \mbox{Same}(\pgt{x},\pgt{y},\pgt{a},\pgt{b},x_{min},x_{max}) \\
G_3 & \pgt{x}'<\pgt{x} \land \pgt{y}'<\pgt{y} \land \mbox{Same}(\pgt{a},\pgt{b},x_{min},x_{max}) \\
\end{array}
$

\vspace{\belowdisplayskip}

The reader may note that the constraints cannot express that $x_{max}$ is \emph{precisely} the maximum
among initial values---which was our intention---but the effect is the same. If for some flow-point we can deduce
the invariant $x_{max}\ge x_j$, then
$x_j$ must be bounded by one of the initial values---since $x_j$ is related to
$x_{max}$ only by paths passing through $x_1,\dots,x_{ar(f_0)}$ at $f_0$. More formally:

\blem
Let $M$ be a rooted multipath of $I(\mathcal{A})$. Suppose that $M$ is satisfiable. Then there is a
satisfying assignment for $M$ such that $\sigma[0,{max}]$ (the assignment of $x_{max}$ in the initial
flow-point) is exactly $\max_{1\le i \le n} \sigma[0,i]$; and $\sigma[0,{min}]$ is exactly $\min_{1\le i \le n} \sigma[0,i]$. 
Such an assignment will be called \emph{tight}.
\elem

\bprf
By assumption, $M$ has a satisfying assignment, say $\sigma'$. We define $\sigma$ to be identical to $\sigma'$
except possibly on the $x_{min}$ and $x_{max}$ variables, whose value we redefine to the values stated in the lemma.
These values satisfy all the constraints in which these variables are involved, and is therefore a satisfying assignment
(all other constraints are satisfied since they are satisfied by $\sigma'$).
\eprf


The next step will be to compute the stable program $S(I(\mathcal{A}))$.
Then we proceed to identifying bounded variables.
To see why stabilization is necessary, consider again the program in Figure~\ref{fig:bounding}.
It is easy to see that at point \pgt{W}, there is no invariant that bounds one of the variables (or both) in terms of
the input values. The closest we might come is to establish a disjunctive invariant of the form ``either the
value of  \pgt{x} or the value of  \pgt{y} is bounded by the input  \pgt{b},'' but such an invariant is not useful
for our approach, as will be seen below.
Stabilization solves this problem: it splits the flow-point $W$ into two points, $W_1$ and $W_2$, representing
the possible cases ($\pgt{x} < \pgt{b}$ and $\pgt{y} < \pgt{b}$). The system $S(I(\mathcal{A}))$ appears in
Figure~\ref{fig:stabilizedexample}.

\begin{fig0}{Example continued: the instrumented and stable system.
The constraints of $I_{f_0}$ are propagated by the stabilization process
to the other flow points reachable from it.
For readability, we do not include the state invariants among the constraints of the adjacent transitions, though they should be there.}{t!}{fig:stabilizedexample}
\begin{gather*}
\xymatrix@R=20pt{
  & *++[o][F]{f_0}\ar_{G_0}[d] & \\
  & *++[o][F]{I}\ar_{G_1}[dl]\ar^{G_2}[dr] & \\
  *++[o][F]{W_1}\ar@(l,d)[]_{G_3}  & & *++[o][F]{W_2}\ar@(l,d)[]_{G_4} \\  
}
\\
\def\arraystretch{1.2}
\begin{array}{@{}l@{\ :\ }ll}
I_{f_0} & x_{max} \ge \pgt{x} \land  x_{max} \ge \pgt{y} \land x_{max} \ge \pgt{a} \land x_{max} \ge \pgt{b}  \ \land\\
\multicolumn{2}{l}{\hspace{2em} x_{min} \le \pgt{x} \land x_{min} \le \pgt{y} \land x_{min} \le \pgt{a} \land x_{min} \le \pgt{b}}\\
G_0 &  \mbox{Same}(\pgt{x},\pgt{y},\pgt{a},\pgt{b},x_{min},x_{max}) \\
\multicolumn{2}{l}{\!\!I_I\;\;  = I_{f_0}} \\
G_1 & \pgt{x}'<\pgt{b} \land \mbox{Same}(\pgt{a},\pgt{b},x_{min},x_{max}) \\
G_2 & \pgt{y}'<\pgt{b} \land \mbox{Same}(\pgt{a},\pgt{b},x_{min},x_{max}) \\
I_{W_1} & \pgt{x}<\pgt{b} \land \pgt{x}>\pgt{a} \land \pgt{y}>\pgt{a} \land
x_{max} \ge \pgt{x} \land  x_{max} \ge \pgt{a} \land x_{max} \ge \pgt{b} \ \land \\
\multicolumn{2}{l}{\hspace{2em}  
  x_{min} < \pgt{x} \land x_{min} < \pgt{y} \land x_{min} \le \pgt{a} \land x_{min} \le \pgt{b}}\\
I_{W_2} & \pgt{y}<\pgt{b} \land \pgt{x}>\pgt{a} \land \pgt{y}>\pgt{a} \land 
x_{max} \ge \pgt{x} \land  x_{max} \ge \pgt{a} \land x_{max} \ge \pgt{b} \ \land \\
\multicolumn{2}{l}{\hspace{2em}  
  x_{min} < \pgt{x} \land x_{min} < \pgt{y} \land x_{min} \le \pgt{a} \land x_{min} \le \pgt{b}}\\
G_3 & \pgt{x}'<\pgt{x} \land \pgt{y}'<\pgt{y} \land \mbox{Same}(\pgt{a},\pgt{b},x_{min},x_{max}) \\
G_4 & \pgt{x}'<\pgt{x} \land \pgt{y}'<\pgt{y} \land \mbox{Same}(\pgt{a},\pgt{b},x_{min},x_{max})  \\
\end{array}
\end{gather*}
\end{fig0}

\bdfn \label{def:B}
For every flow-point $f$ of $S(I(\mathcal{A}))$, let $$B(f) = \{ j \mid (x_j \succ^b x_{min}), (x_{max}\succ^d x_j)
\in I_f  \text{ for some $b,d$} \}\,. $$
We call $B(f)$ the set of  \emph{bounded variables} at $f$.
\edfn


\subsection{Deciding Bounded Termination}


We next provide a decision algorithm for bounded termination.
The idea, in a nutshell:  ignore all non-bounded variables, and check for termination.
The example given earlier illustrates that the assumption of stability in
Definition~\ref{def:B} is crucial for the correctness of this algorithm, which is why it is applied to $\ourmcs$.
To formalize the idea of checking for termination while looking only at the bounded variables (Definition~\ref{def:B}),
we define, more generally, satisfaction of multipaths, and termination, under restriction to any given choice of variables per flow-point.

\bdfn
If $M$ is an $\mathcal A$-multipath, a \emph{partial assignment} for $M$
is an assignment $\sigma$ to the variables of $M$ that assigns integers to 
some
variables and the special value $\bot$ to all others. This value satisfies any constraint (including $\bot>\bot$).
Given a function $C$ that associates to each $f\in \F{A}$ a subset of $\{1,\dots,ar(f)\}$ (representing a choice of some variables),
$\sigma$ is a $C$-restricted assignment if the variables indicated by $C$ are those that receive integer values.
Multipath $M$ is called \emph{satisfiable on $C$} if it has a $C$-restricted assignment that satisfies all constraints
(note that only constraints among the chosen variables really matter).
\edfn

\bdfn
For a choice function $C$ as above,
We say that $\ourmcs$ \emph{terminates on $C$} if there is no infinite, rooted multipath $M$ which is satisfiable
on $C$.
\edfn

\pagebreak[4]
\begin{alg} (Bounded Termination) \label{alg-BT}
Input: {\MCTS} $\mathcal A$.
\end{alg}
\be
\item 
Build $\ourmcs$.
\item
Perform a decision procedure for termination on $\ourmcs$, taking only bounded variables into account.
\item 
Return the result of the termination procedure.
\ee
Clearly, the algorithm checks for termination on $B$ of $\ourmcs$. We next prove that this is a sufficient and necessary condition
for bounded termination.


\bthm \label{thm:decision-pi-soundness}
If $\ourmcs$ terminates on $B$, then 
$\cal A$ bounded-terminates. Moreover, 
let $\maxi$ and $\mini$ be the maximum and minimum values among the variables of the initial state.
The height of the initial state is 
$O((\maxi - \mini)^n)$, where the constant factor depends on the size of $\ourmcs$,
and $n$ is the maximum of the flow-point arities in $\cal A$.
\ethm

\bprf
The fact that $\cal A$ is terminating follows immediately as termination on $B$ is a stronger notion than termination
(if an infinite multipath is not satisfiable on $B$, it is not satisfiable).
To justify bounded termination, consider any rooted $\ourmcs$-multipath, and a satisfying, tight
$B$-restricted assignment. All bounded variables will be assigned values between $\mini$ and
$\maxi$. The variables $x_{max}$ and $x_{min}$ are constant throughout.
Thus there are at most $m(\maxi-\mini)^n$ different states that can potentially appear in a satisfying assignment to this multipath, where
$m$ is the number of flow-points in $\ourmcs$.
There can be no repeated states, since otherwise one can use an obvious ``cut and paste'' argument
and exhibit an infinite  multipath satisfiable on $B$.
We conclude that the height of the initial state is bounded by $m(\maxi-\mini)^n$.
\eprf

To prove completeness, we use the following fact.

\blem  {\upshape \cite{BA:mcInts}} \label{lem:loop} 
If $\cal A$ is a non-terminating {\MCTS} with initial point $f_0$,
there is a flow-point $f$, a cyclic multipath $L:f\leadsto f$,
and a rooted multipath $H:f_0\leadsto f$,
such that $H L^\omega$ ($H$ followed by an infinite sequence of $L$'s) is satisfiable.
\footnote{
Actually, the corresponding lemma in~\cite{BA:mcInts} does not consider rooted termination and
therefore neglects the stem $H$. But if the termination test is modified to test for rooted
termination (so that only reachable cycles are considered), the lemma stated here ensues.
}
\elem

And we add a new lemma.

\blem[Assignment Extension] \label{lem:extendass}
Let $M$ be a finite multipath in $\ourmcs$
 and $\sigma$ a $B$-restricted assignment for $M$. 
It is possible to extend $\sigma$ to an assignment $\sigma'$ that satisfies $M$.
\elem

Extending $\sigma$ simply means assigning integer values to the variables left undefined
($\bot$) by $\sigma$, which are the non-bounded variables appearing in $M$.

\bprf
We treat $M$ as a directed graph, with arcs weighted by $0$ for a non-strict arc and $-1$ for
a strict one. Since $M$ is satisfiable (Lemma~\ref{lem:stable=satisfiable}), there is no negative-weight
cycle;  hence, for all nodes $u$ and $v$, if $v$ is reachable from $u$, there is a minimum-weight path from $u$ to $v$.
We define the \emph{minimum-weight distance} $\delta(u,v)$ to be the weight of such a path,
 or $+\infty$ if $v$ is unreachable from $u$.
 
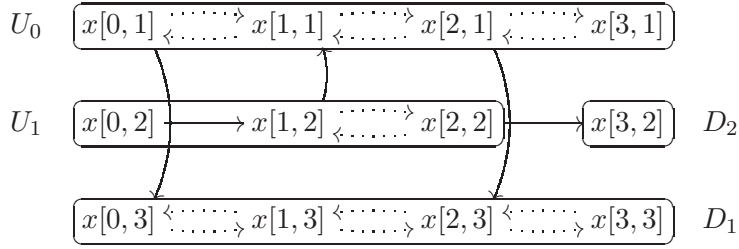
\begin{fig0}{An example to the construction in the proof of the Assignment Extension lemma. To simplify the drawing,
$x_{min}$ and $x_{max}$ are not shown; assume that they exist and 
that all the constraints involving them are $x_{min} \le x_0 \le x_{max}$,
causing $U_0$ to be as shown.}{t}{fig:UDsets}
\SelectTips{cm}{}
$$\xymatrix@R=20pt@C=30pt{
U_0 \kern -30pt &  x[0,1]\ar@<2ex>@/^10pt/@{->}[dd]\ar@/^5pt/@{.>}[r] & x[1,1]  \ar@/^5pt/@{.>}[l] \ar@/^5pt/@{.>}[r] & x[2,1]\ar@<2ex>@/^10pt/@{->}[dd]\ar@/^5pt/@{.>}[r]\ar@/^5pt/@{.>}[l] & x[3,1] \ar@/^5pt/@{.>}[l]\\
\save "1,2"."1,5"*[F-:<3pt>]\frm{}
\restore
U_1 \kern -30pt & x[0,2] \ar@{->}[r]  & x[1,2] \ar@/_5pt/@<-2ex>@{->}[u]\ar@/^5pt/@{.>}[r]  & x[2,2]\ar@/^5pt/@{.>}[l]\ar@{->}[r]
  & *+[F-:<3pt>]{x[3,2]} & \kern -30pt {D_2} \\
\save "2,2"."2,4"*[F-:<3pt>]\frm{}
\restore
 & x[0,3]\ar@/_4pt/@{.>}[r] & x[1,3]\ar@/_4pt/@{.>}[l]\ar@/_4pt/@{.>}[r]
  & x[2,3]\ar@/_4pt/@{.>}[l] \ar@/_4pt/@{.>}[r]& x[3,3]\ar@/_4pt/@{.>}[l] & \kern -30pt {D_1} 
\save "3,2"."3,5"*[F-:<3pt>]\frm{}
\restore
}$$
\end{fig0}
 
We define sets of nodes $U_0,D_1,U_1,D_2,U_2\dots$ as follows:
\begin{itemize}
\item $U_0$ consists of all nodes which represent \emph{bounded} variables.
\item For all $i\ge 0$, let $P^D_{i+1} = U_0 \cup D_1 \cup \dots \cup U_i$; then,
$D_{i+1}$ is the set of nodes $v\notin P^D_{i+1}$
such that  $u\leadsto v$ for some $u\in P^D_{i+1}$.
\item For all $i\ge 1$, let $P^U_{i} = U_0 \cup D_1 \cup \dots \cup D_i$; then,
$U_{i}$ is the set of nodes $u\notin P^U_{i}$
such that  $u\leadsto v$ for some $v\in P^U_{i}$.
\end{itemize}

See Figure~\ref{fig:UDsets} for an example.

We extend $\sigma$ from the nodes of $U_0$, on which it is initially defined, to nodes of every
set $D_i$ and $U_i$, inductively.  Suppose that $U_0, D_1, \dots,U_i$ have already been treated;
let $P^D_{i+1}$ be the union of these sets.
Then for every $v\in D_{i+1}$ we extend $\sigma$ by letting 
$$\sigma(v) = \min_{u\in P^D_{i+1}} \{ \sigma(u) + \delta(u,v) \}$$
Note that $\sigma(v)$ is finite since, by definition of $D_{i+1}$, there are nodes in $P^D_{i+1}$ such that
$\delta(u,v)$ is finite, and they are already assigned.

Alternatingly, we extend $\sigma$ to $U_{i}$, assuming that all nodes in $P^U_{i} = U_0\cup D_1\cup\dots
\cup D_i$ have been assigned. For $u\in U_{i}$ we let
$$\sigma(u) = \max_{v\in P^U_{i}} \{ \sigma(v) - \delta(u,v) \}$$
As above, $\sigma(v)$ is well-defined and finite.

We claim that the assignments to $\sigma$ are consistent with the constraints in $M$. To prove this,
consider an assignment to $v\in D_{i+1}$. There are three possible types of constraints involving $v$ and another assigned variable:

(1) A constraint $v\succ^b v'$ with both $v$ and $v'$ in $D_{i+1}$. Thus, both are reachable from $P^D_{i+1}$,
and, by the definition of $\delta$, we have for all $u\in P^D_{i+1}$, $\delta(u,v') \le \delta(u,v) + b $.
In particular, choose $u_L\in P^D_{i+1}$ such that $\sigma(u_L) + \delta(u_L,v)$ is minimum (and, hence,
this is the value assigned to $v$); then
$$\sigma(u_L) + \delta(u_L,v')  \le  \sigma(u_L) + \delta(u_L,v) + b $$
so our definition of $\sigma(v)$ and $\sigma(v')$ satisfies
$$\sigma(v') \le \sigma(v) + b$$
and the constraint is satisfied.

(2) A constraint $v\succ^b v'$ with $v$ in $D_{i+1}$ and $v'$ in $P^D_{i+1}$.
Here the case $i=0$ is special. So consider first $i>0$. 
By examining the definition of the sets, the reader may verify that 
$P^D_{i+1}$ is closed under reverse-reachability. Hence, our assumptions imply $v\in P^D_{i+1}$,
and $v\in D_{i+1}$ is impossible.
Next, let $i=0$; so $v$ is in $D_1$ and $v'$ in $U_0$ which is the set 
of bounded variables. 
By the definition of $D_1$,
there is a $U_0$ variable which upper-bounds $v$, while
$v'$ lower-bounds it, so $v$ too is a bounded variable and cannot be in $D_1$.

(3) A constraint $v'\succ^b v$ with $v$ in $D_{i+1}$ and $v'$ in $P^D_{i+1}$.
Clearly, 
$$\min_{u\in P^D_{i+1}} \{ \sigma(u) + \delta(u,v) \} \le \sigma(v') + b$$
so this constraint will be satisfied.

A similar case analysis justifies the assignments in $U_{i}$.

Finally, there may remain nodes not in any of the above sets. These nodes are not connected to
any node already assigned. So an assignment may be chosen for them freely, only having to satisfy
relations among themselves, which is possible since $M$ is satisfiable.
\eprf


\bthm [Completeness] \label{thm:decision-pi-completeness}
A {\MCTS} $\cal A$ bounded-terminates only if $\ourmcs$ terminates on $B$.
\ethm

\bprf
Suppose that $\ourmcs$ does not terminate on $B$. We claim that $\cal A$ is not bounded-terminating,
which means that there is 
an initial state which can be followed by runs arbitrarily long.

We use Lemma~\ref{lem:loop}. It provides us with
a cyclic multipath $L$ and a rooted multipath $H$, such that $HL^\omega$ is satisfiable on $B$, say by $\sigma$.
For every $p\ge 0$, multipath $HL^p$ is satisfied on $B$ by the corresponding restriction of $\sigma$.
Since it is a finite multipath in a stable system, by Lemmas~\ref{lem:stable=satisfiable} and
\ref{lem:extendass} we can extend $\sigma$ to a complete assignment $\sigma_p$ that satisfies $HL^p$.

Next, we note that all the variables of the initial point $f_0$ of $I(\mathcal{A})$ are clearly bounded,
which means that $\sigma$ valuates them. So, all the assignments $\sigma_p$ agree on the initial state.
This concludes the proof.
\eprf

Finally we consider the complexity of the decision problem.

\bthm\label{thm:decision-pi-complexity}
Deciding whether a {\MCTS} $\cal A$ bounded-terminates is PSPACE-complete (and is PSPACE-hard
even for stable systems that have a single flow-point).
\ethm

\bprf

\emph{Upper bound}.  The algorithm as described constructs $\ourmcs$, which is a difficulty
since stabilization may,
in general, increase the size of a {\MCTS} exponentially. However, as shown in~\cite{BA:mcs}, it is
possible to implement the decision procedure for termination (or, more precisely, for non-termination)
as a non-deterministic PSPACE algorithm.
The problem is then in PSPACE thanks to Savitch's theorem.
The trick is to use full elaboration (which yields a stable system).

Given the {\MCTS}  program  $\mathcal{A}$, our algorithm 
constructs flow-points and transitions of the elaborated system $\elabmcs=E(I(\mathcal{A}))$ on the fly.  
First, it non-deterministically selects an initial flow-point; then it walks through $\elabmcs$ to find a reachable flow-point $f$ that it
 guesses will start the loop (the part denoted above by $L$). From that point on, it maintains a summary of the
multipath traversed (namely the collapsed multipath $\overline M$, as in Definition~\ref{def:collapse}).
It proceeds with the random walk through $\elabmcs$ until
a counter-example to bounded termination (a cyclic multipath which is not terminating on $B$) has been found
(note that one can determine the termination property just from $\overline M$; this is the basis for the 
``closure algorithm" for {\MCTS} termination~\cite{BA:mcInts}).

\emph{Lower bound}. We reduce from the SCT problem~\cite{leejonesbenamram01}, a simple case
of {\MCTS} termination.  In an SCT instance, the only type of constraints that appear in the input
is $x_i \succ x_j'$ (a source variable bounds a target variable). 
Furthermore, we restrict the problem to singleton control-flow graphs (that is, there is a single flow-point).
This restricted problem is known to be PSPACE-hard~\cite{Ben-Amram:ranking}. 

Let $\cal S$ be an SCT instance, with a single flow-point and with $n$ variables. Add variables $x_b$ (bottom) and $x_t$ (top),
and the constraints: $x_t \ge x_i \ge x_b$, for every $i$. To every
transition, add the constraints $x_b = x_b'$ and $x_t = x_t'$, plus the constraints of the adjacent
flow-points.
We claim that the resulting system, $\cal A$, bounded-terminates if and only if $\cal S$ terminates.
Indeed, it is obvious that we made all the variables of $\cal S$ bounded. So if $\cal S$ terminates,
$\cal A$ satisfies the condition for bounded-termination. For the other direction, suppose that
$\cal S$ does not terminate:  \cite{Codish-et-al:05} shows that in such a case, there is a loop
in ${\mathcal T}_{\mathcal S}$. That is, there is a run which reaches
a certain state $s$ and then repeats forever a certain finite run from $s$ to $s$. 
Such a run, though infinite, only includes a finite set of integer values. By setting the initial value
of $x_b$ to the  minimum of these values, and that of $x_t$ to their maximum, we obtain an infinite run
of $\cal A$. To conclude, if $\cal S$ is non-terminating, so is $\cal A$.

We have thus reduced the single-flow-point SCT problem, known to be PSPACE-hard, to bounded termination.
It is easy to verify that the {\MCTS} created is also stable, concluding the proof of the theorem.
\eprf

\section{Significance for Concrete Programs}
\label{sec:programs}

\MCTS{}s may be considered as an abstract computational model and its analysis as a goal in itself,
which is interesting since such systems, despite the relative simplicity of the constraints, may exhibit
a complex behaviour. However, we would like to promote the view that such systems are useful
as an abstraction of concrete programs, to facilitate their analysis.

In this section we consider the question:
What does the fact that a {\MCTS} has polynomially bounded height tell us about the program it 
represents?  
We discuss this question in three settings, which we first present informally; secondly, we give a formal
example using a toy programming language defined for this sake only; and finally relate our discussion
to the implementation of two published analysis tool which use the \emph{abstract and conquer} approach.

\subsection{The three settings, informally}
\label{sec:programs1}

\paragraph{Flat imperative programs}
We first consider imperative programs without any
procedure calls (hence ``flat").
Figures~\ref{fig:ex:A} and \ref{fig:ex:quadratic1} are examples of
flat imperative programs abstracted in the natural way.
The control-flow graph corresponds to the flow-chart of the program;
transitions correspond to program instructions, or---more effectively---basic blocks.
Often, the assumption is that such a block takes a constant time to execute.

In this setting, the height of the transition system
represents the time complexity of the program. In terms of complexity classes, this allows us to
identify a program as polynomial-time in the selected input parameters.
When basic blocks have associated costs which are not uniform, the Reachability Bound analysis~\cite{Gulwani-PLDI10}
may allow for infering a bound on the cost of a computation based on the formula
$\sum_f \text{cost}(f)\times T_f$ where $f$ ranges over flow-points and $T_f$ is a reachability bound
for $f$. 

In practice, the control-flow graph of the program may be transformed during abstraction. Suppose that
we select a set of \emph{cut points} in the program's flow-chart such that any cycle must traverse
a cut point, and the program entry is a cut point.
Any such set of cut points may be chosen as the set of flow-points as long as any (finite) path
between two cut points is represented by an abstract transition. The conclusions on the
concrete program's complexity remain valid.

For programs that contain procedure calls, but not recursion, a bottom-up analysis may be applicable.
The results of analyzing a procedure \pgt{p} will be plugged into the summation for its caller, using
reachability bounds, as shown above (Albert et al.~\cite{AAGP:jar2010} also describe a bottom-up process,
however their analysis is not based on the RB approach).

\paragraph{Pure-functional programs}
Lee et al.~\cite{leejonesbenamram01} showed the simplest way in which a (first-order, eager) pure-functional
 program may be abstracted.
The control-flow graph is the \emph{call-graph} of the program;
flow-points are function names and
transitions correspond to functions calls. 
Hence, every \emph{call chain}
 of the program corresponds to a particular run of the transition system.

It should be clear that in this setting, the height of the transition system
represents the \emph{stack height} of the program. This is a resource of practical
importance in itself. What can we infer in terms of the traditional resources, space
and time?  
The pure functionality
suggests that there is no iteration but recursion, so it may be possible to bound the
execution time of a function body, or the space it consumes, outside any calls it performs;
often, this bound will be a constant.
If functions cannot allocate ``heap space'' at all, the stack height
corresponds to space usage.  If the functions can allocate space outside the stack,
exponential space may be consumed for a polynomial stack height.
Because the call tree is a tree
of bounded degree, we obtain an exponential time bound (that is, a constant to a polynomial power).

In terms of complexity classes, we may conclude that the program is polynomial-space or only the
weaker result that it has the class EXPTIME.

Also in this setting, we note that the abstraction may create the CFG in different ways, which are
sometimes useful. For example, Manolios and Vroon~\cite{MV-cav06} chose call sites to be flow-points rather than function names.

\subsection{Analysing a simple programming language}
	
We demonstrate the ideas more formally by defining three variants
of a simple (but Turing complete) functional programming language ${\spfl}$
 and a simple-minded, conservative abstraction
$\abst$ mapping ${\spfl}$ programs to {\MCTS}s.
Since the language has functional style, imperative programs are represented by tail recursion.

The syntax of ${\spfl}$ is defined in Table~\ref{fig:syntax}, and further explained below. Semantically,
$\spfl$ programs operate on strings over a finite alphabet $\Sigma=\{0,1,\dots\}$.
The expression
$a{:}x$, where $a\in \Sigma$, evaluates to $a$ followed by the value of $x$.

A program is a collection of definitions which leaves no 
undefined identifiers. A function is defined by a set of definitional
patterns. To avoid ambiguity, a first-match disambiguation rule (as in ML) can be used.
If there is no match, the program halts. A wildcard ``?'' can be introduced 
in patterns as syntactic sugar.
For simplicity, all functions have the same arity $n$.
A function $\finit$ is indicated as the entry point. 

\begin{table}[t]
\setlength{\arraycolsep}{1mm}
$$\begin{array}{llrlll}
{\it Prg} & \ni & \pgt{p} & ::= & D_1\ \ldots\ D_N\\
{\it Dfn} & \ni & D_i & ::= & f(\pi_1,\ldots,\pi_n) = e \\
{\it Expr} & \ni & e & ::= & \alpha_1' \\
&&&&\qquad \mbox{(simple expression)}\\
&&& \multicolumn{2}{l}{|\  g(\alpha_1,\ldots,\alpha_n) } \\
&&&&\qquad \mbox{(tail-recursive expression)}\\
&&& \multicolumn{2}{l}{|\  g_1(\alpha_1,\ldots,\alpha_n) \:\pgt{?}\ 
g_2(\alpha'_1,\ldots,\alpha'_n), g_3(\alpha''_1,\ldots,\alpha''_n) } \\
&&&&\qquad \mbox{(conditional expression)}\\
&&& \multicolumn{2}{l}{|\  \pgt{let}\ y = g_1(\alpha_1,\ldots,\alpha_n) 
 \ \pgt{in}\ g_2(\beta_1,\ldots,\beta_n) } \\
&&&&\qquad \mbox{(nested expression)}\\
{\it Pat} & \ni & \pi_i & ::= & \eps \mid x_i \mid a{:}x_i \\
&&&&\mbox{(parameter pattern)}\\
{\it APat} & \ni & \alpha_i' & ::= & \eps \mid a \mid x_j \mid a{:}x_j \mid b{:}a{:}x_j \\
&&&&\mbox{(actual parameter)}\\
{\it APat}' & \ni & \beta_i & ::= & \alpha_i' \mid y \\
&&&&\mbox{(extended actual parameter)}\\[1ex]
 \multicolumn{5}{l}{a,b\in\Sigma \qquad i,j\in\{1,\dots,n\}} \\
\end{array}$$
\caption{Syntax of {\spfl}}
\label{fig:syntax}
\end{table}

\bex
Here is a short {\spfl} program that tests two strings for equality,
where $\Sigma=\{0,1\}$.
For some complication, it occasionally swaps its arguments.
We use the first-match rule for pattern matching.
\begin{align*}
& f(\eps, \eps) = 1 \\
& f(0{:}x_1, 0{:}x_2) = f(x_1, x_2) \\
& f(1{:}x_1, 1{:}x_2) = f(x_2, x_1) \\
& f(x_1, x_2) = \eps 
\end{align*}
\eex

\bex
The next program has exponential time and space complexity.
\begin{align*}
& f(\eps, x_2, \eps) = x_2 \\
& f(x_1, x_2, 1:x_3) = f(x_1, 1:x_2, x_3) \\
& f(1{:}x_1, x_2, \eps) = f(x_1,x_2, x_1) 
\end{align*}
\eex

The specification of function bodies and their return values differs in the
three language variants:
\begin{description}\setlength{\labelsep}{4mm}
\item[${\spfl}_1$] Allows only the simple and the tail-recursive 
expressions as function bodies. Hence, it represents imperative programs.
The return value of functions is $\Sigma^*$. 

\item[${\spfl}_2$] Allows, in addition, the conditional expression.
The condition $g_1(\dots)$ is evaluated first; if it is
a non-empty string, the value of the expression is obtained by evaluating $g_2(\dots)$,
and otherwise, $g_3(\dots)$.

\item[${\spfl}_3$] Also includes the nested (``let") expression.
\end{description}

\bdfn
Abstraction $\abst$ maps an {\spfl} program to a {\MCTS} as follows:
the flow-point set $F$ is the set of defined functions. 
There is an abstract transition $G:f\to g$ for every call expression
$g(\rho_1,\ldots,\rho_n)$ in a definition $f(\pi_1,\ldots,\pi_n) =
\dots$; a relation among $x_i$ and $x_j'$ is included in $G$, dependent on the
patterns $\pi_i$ and $\rho_j$, 
as specified in the following table (the cases missing in the table contribute no constraint).

\[
\begin{array}{cccl}
\pi_i & \rho_j & \text{relation} \\ 
\hline
\eps    &  \eps & = \\
x_i    &  \eps  &  \ge \\
x_i    &  x_i  & = \\
a{:}x_i    &  \eps  \mid x_i  & >  \\
a{:}x_i    &  a'   & \ge \\
a{:}x_i    &  a'{:}x_i  & =  \\
\end{array}
\]
\edfn
\noindent 
We can now state our observations in this formal setting.

We assume a typical RAM implementation of $\spfl$, using a stack for function calls,
and a heap memory to keep the strings,
which are implemented as linked lists, so that removing or adding an element at the front takes
constant time and space. We also assume immediate garbage collection so that garbage does not
accumulate (this is easy for such  a language, e.g., by reference counting).

\begin{claim} \label{claim-spfl}
If $\abst(\pgt{P})$ satisfies bounded termination, where \pgt{P}
is an ${\spfl}_i$ program, then, for all $i$, the stack height is 
polynomially bounded in the size of the input strings. For
$i=1$, the program runs in polynomial time; for $i=2$, its space usage
is polynomial;
and for $i=3$, its running time is bounded by $2^{poly(n)}$.
\end{claim}

A formal proof of this claim is skipped as it is uninteresting
and tedious (demanding a formalization of semantics and complexity,
currently left informal).  The time bound in the case of $\spfl_1$ is straightforward
and that of $\spfl_3$ follows almost as easily since the height of the recursion tree is polynomial.
As to the space bound for $\spfl_2$, note that a branch in the recursion tree only occurs in this language
when a conditional is evaluated, and that heap space allocated by the evaluation of the condition ($g_1$)
can be discarded once it is determined whether the return value is $\eps$ or not. Thus, for the purpose
of bounding the space, it is possible to consider the stack height.

Note that our language is Turing-complete. It is possible to extend Claim~\ref{claim-spfl}
to a proposition of \emph{class capture}: every decision problem in PTIME (resp., PSPACE,
EXPTIME) may be represented by an ${\spfl}_1$ (resp.,  ${\spfl}_2$, ${\spfl}_3$) program.
We find that this result is of little consequence to the main goals of our work, and have decided
to omit the proof.

\subsection{A discussion of two analyzers for real-world languages}
	
We compare our informal statements at the beginning of this section to
the way abstraction is used in the WTC project by Alias et al.~\cite{ADFG:2010}
and the COSTA project by Albert et al.~\cite{Albert-et-al:FSAD:2009,Albert-et-al:TCS:2011}.
As described in Section~\ref{sec:intro}, both works use a constraint language richer
than monotonicity constraints, but this issue is independent of the current discussion
(it may affect \emph{precision} of the abstraction---see the next section).

In~\cite{ADFG:2010}, C language programs are abstracted to affine constraint transition systems.
They have implemented two forms of abstraction. One represents a basic block as a transition,
another only places a flow-point at a loop header and expands the loop body so that every path
through the loop is abstracted to one abstract transition. This means that exponentially more 
transitions may be generated, but the abstraction will be more precise. In both cases, our
informal description for ``flat imperative programs'' applies.

In~\cite{Albert-et-al:FSAD:2009,Albert-et-al:TCS:2011}, Java Bytecode programs are abstracted to transition systems which
express a sequential transition (from a block in the flow-chart to the next) and a procedure call
in essentially the same way. Thus a sequential computation is treated as tail recursion---much like
in our toy language. The analysis described in~\cite{AAGP:jar2010} distinguishes the case of
tail recursion from the case where a recursion tree is involved and an exponential bound may
result.
This is again similar to the framework we have described. Their abstract programs are annotated
with cost expressions, used in computing a closed formula for a cost bound. As stated earlier,
in our framework this may require the computation of reachability bounds and a (symbolic) summation,
and possibly also another static analysis to bound the cost expressions in terms of input parameters.

There are other tools that translate real-world languages to some kind of contraint transition
systems, for example~\cite{SpotoMP09} analyze Java Bytecode and \cite{MV-cav06} analyze the
ACL2 programming language, both for the purpose of termination analysis. Since the
correspondence of the abstract program to the concrete one is still essentially as in our discussion,
we conclude that the generated abstract programs could be used, perhaps with some adaptation, for cost analysis as well.

\subsection{Reflections on effective abstraction}
\label{sec:realworld}



Both of the tools we cited in the last section use a more expressive abstraction---an affine-constraint
CTS (also known as a CTS with polyhedral constraints). This constraint language is strictly more
expressive, as monotonicity constraints form a simple special case of affine constraints.
So there is reason to fear that by abstracting a program to a \MCTS we might lose crucial
information.
We would like to argue that this consideration
should not discourage researchers from employing this abstraction.

One reason for our optimism is the existing empirical evidence for the effectiveness of the size-change
technique in termination analysis \cite{LS:97,CT:99,TG:05,MV-cav06,BC:08:TACAS,SpotoMP09,KraussHeller09,CGBFG:11:TPLP}.
As shown in our theoretical sections, the complexity analysis is a refinement of termination analysis
and reuses its methods.
Nonetheless, we argue that for bounded termination analysis, it is necessary to transfer more
information to the \MCTS than one does for termination,
 in particular if one wants to analyze it as a stand-alone abstract program.
The main reason is the necessity for \emph{bounding variables}. Consider Program~2
in Figure~\ref{fig:ex:quadratic1} on Page~\pageref{fig:ex:quadratic1}.
If the initial assignment is changed from \texttt{i=N} to \texttt{i = 2*N}, and the abstract variables still
correspond to the program variables in a one-to-one fashion, we will lose the bound on
\pgt{i} in terms of \pgt{N}, since it is not a monotonicity constraint.  Note that this relation is not necessary
for the termination proof, but is crucial for deducing \emph{bounded} termination.

We think that this problem may be mitigated by the use of an auxiliary bound analysis, one which
attempts to bound expressions in the program in terms of the designated input variables. Such an
analysis can be performed by, for example, polyhedral analysis \cite{CousotHalbwachs:1978} or one
of its many variants. When an expression \textit{exp} is found to be bounded by a bound
$B_{exp}$ in terms of the input, an abstract variable representing $B_{exp}$  may be added to the
abstraction. In order to avoid combinatorial explosion, one may decide to add such variables only
when necessary for changing an unbounded variable in the \MCTS into a bounded one; one may also
opt to keep only a representative of the maximum among such expressions, in the same way we used
$x_{max}$ in Section~\ref{sec:mainThm}. Note that if we have an analysis that (unlike polyhedral
analysis) may ascertain a non-polynomial
bound on  \textit{exp} we may end up with complexity bounds that are polynomial functions of that bound,
hence possibly non-polynomial as a function of the input parameters.

We also invite the reader to note that \MCTS{}s can capture rather complex behaviours. The examples in
the next section illustrate a few.  This should be at least a reason to consider the model interesting.


\section{Additional Examples}
\label{sec:examples}

To illustrate the variety of loop structures that can be represented and analysed,
we have selected a few examples, shown in this section as C program fragments;
see also examples on pages \pageref{fig:ex:A}, \pageref{fig:ex:quadratic1}, \pageref{fig:bounding}.
In all these examples, it is pretty simple to verify that the associated constraints systems are indeed
bounded terminating.

\bex 
This is a quadratic-time example (similar to Figure~\ref{fig:ex:quadratic1},
but counting up rather than down), from~\cite{SPEED-POPL09}, where it is analysed
by means of counter instrumentation and bounding. 

\begin{program}
SimpleMultipleDep(int n, int m) \{
    x = 0; 	y = 0;
    while (x < n)
        if (y < m) y++;
        else \{ y = 0; x++; \}
\}
\end{program}

\noindent
Here is its MC representation:

$$\raisebox{30pt}{
$
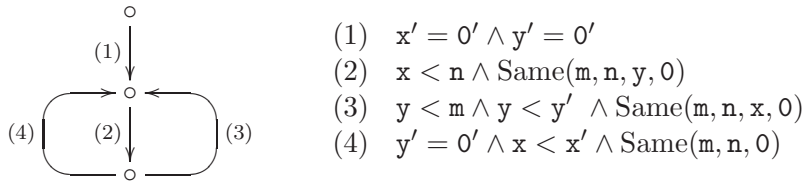
\xymatrix@R=20pt{
 & \circ\ar[d]_{(1)}
& \\
 & \circ\ar[d]_{(2)}
& \\
  & \circ\ar`r[ur] `[u]_{(3)} [u] \ar`l[ul] `[u]^{(4)} [u]& 
}$} \qquad 
\begin{array}{ll}
(1) & \pgt{x}'=\pgt{0}' \land \pgt{y}'=\pgt{0}' \\
(2) & \pgt{x}<\pgt{n} \land \mbox{Same}(\pgt{m},\pgt{n},\pgt{y},\pgt{0}) \\
(3) & \pgt{y}<\pgt{m} \land \pgt{y}<\pgt{y}' \ \land \mbox{Same}(\pgt{m},\pgt{n},\pgt{x},\pgt{0}) \\
(4) & \pgt{y}' = \pgt{0}' \land \pgt{x}<\pgt{x}'  \land \mbox{Same}(\pgt{m},\pgt{n},\pgt{0})
\end{array}
$$
\eex

\bex 
The next example is from~\cite{SPEED-POPL09}. They explain that their algorithm
does not handle it because of the lack of path-sensitive information.  Alias et al.~report
in~\cite{ADFG:web} that their tool solved this instance. 

\begin{program}
void pathSensitive2(int n, int b, int x) \{
  int t;
  if (b>=1) t=1; else t = -1;
  while (x<=n) \{
      if (b>=1)
           x=x+t;
      else 
           x=x-t;
  \}
\}
\end{program}

\noindent
In its MC representation, we represent the effect of addition and subtraction disjunctively:
for example, we use the knowledge that \pgt{x = x+t} is a command that increases \pgt{x}
if \pgt{t} is positive, decreases \pgt{x} if \pgt{t} is negative, etc. Thus we have three MCs for each
command of this form. In this particular program, two of those represent transitions that will never
be taken in an actual run, but we do not assume our ``front end" to do such an analysis.

$$\raisebox{60pt}{
$
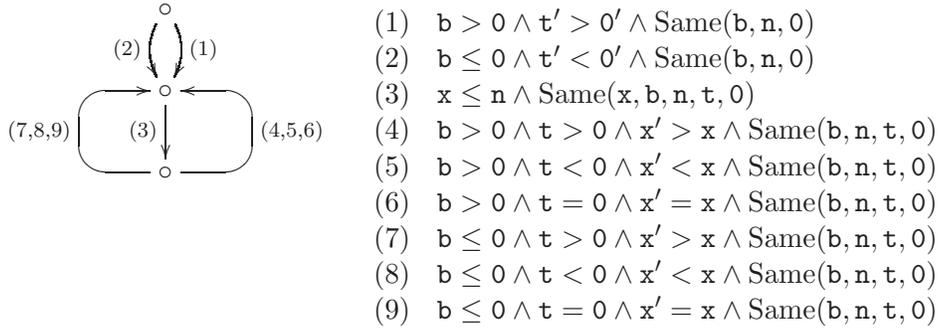
\xymatrix@R=20pt{
 & \circ\ar@/^/[d]^{(1)}\ar@/_/[d]_{(2)}
& \\
 & \circ\ar[d]_{(3)}
& \\
  & \circ\ar`r[ur] `[u]_{(4,5,6)} [u] \ar`l[ul] `[u]^{(7,8,9)} [u]& 
}$} \quad 
\begin{array}{ll}
(1) & \pgt{b}>\pgt{0} \land \pgt{t}'>\pgt{0}' \land \mbox{Same}(\pgt{b},\pgt{n},\pgt{0})\\
(2) & \pgt{b}\le\pgt{0} \land \pgt{t}'<\pgt{0}' \land \mbox{Same}(\pgt{b},\pgt{n},\pgt{0})\\
(3) & \pgt{x} \le \pgt{n} \land \mbox{Same}(\pgt{x},\pgt{b},\pgt{n},\pgt{t},\pgt{0}) \\
(4) & \pgt{b} > \pgt{0} \land \pgt{t} > \pgt{0} \land \pgt{x}'>\pgt{x}  \land \mbox{Same}(\pgt{b},\pgt{n},\pgt{t},\pgt{0}) \\
(5) & \pgt{b} > \pgt{0} \land \pgt{t} < \pgt{0} \land \pgt{x}'<\pgt{x}  \land \mbox{Same}(\pgt{b},\pgt{n},\pgt{t},\pgt{0}) \\
(6) & \pgt{b} > \pgt{0} \land \pgt{t} = \pgt{0} \land \pgt{x}'=\pgt{x}  \land \mbox{Same}(\pgt{b},\pgt{n},\pgt{t},\pgt{0}) \\
(7) & \pgt{b} \le \pgt{0} \land \pgt{t} > \pgt{0} \land \pgt{x}'>\pgt{x}  \land \mbox{Same}(\pgt{b},\pgt{n},\pgt{t},\pgt{0}) \\
(8) & \pgt{b} \le \pgt{0} \land \pgt{t} < \pgt{0} \land \pgt{x}'<\pgt{x}  \land \mbox{Same}(\pgt{b},\pgt{n},\pgt{t},\pgt{0}) \\
(9) & \pgt{b} \le \pgt{0} \land \pgt{t} = \pgt{0} \land \pgt{x}'=\pgt{x}  \land \mbox{Same}(\pgt{b},\pgt{n},\pgt{t},\pgt{0}) \\
\end{array}
$$
\eex

\bex \label{ex:nolex}
The next program does not have a lexicographic-linear global ranking function,
an obstacle for tools that, explicitly or implicitly, require functions of this kind
(this class includes~\cite{ADFG:2010}, by their own description, and
also COSTA, though the fact is implicit---see Section~\ref{sec:rw}.
The class also includes the algorithm
of \cite{SPEED-POPL09},  according to a discussion in~\cite{ADFG:2010}).
We omit the transition system this time, which the reader would be able to create at ease
(for assignments like \pgt{y = y+x} it suffices, in this case, to consider \pgt{y} as being unconstrained,
although a disjunctive representation of the effect, as in the previous example, could be harmlessly
included).

\begin{program}
\noindent
void min(int x, int y) \{
   while (y > 0 && x > 0) \{
      if (x>y) z = y;
        else  z = x;
      if (*)\{ y = y+x; x = z-1; z = y+z \}
       else \{ x = y+x; y = z-1; z = x+z \}
    \}
\}	
\end{program}
\eex

Another instance where lexicographic linear global ranking functions do not suffice is given
in Figure~\ref{fig:bounding} (Page~\pageref{fig:bounding}).

\bex The following example from \cite{Gulwani-PLDI10} shows the weakness of
a straight-forward abstraction to monotonicity constraints.
\begin{program}
\noindent
i = 0;
while (i < n) \{
   j = i + 1;
   while (j < n) \{
     if (A[j])
        j--;  n--;
     j++;
   \}
   i++;
\}   
\end{program}
The problem is that abstracting the effect of the \pgt{if}-block on \pgt{j} to $\pgt{j}' < \pgt{j}$ does not allow a later analysis
to figure out that \pgt{j++} ``undoes'' this decrement.  There are, of course, multiple ways to handle this issue. For example,
one could use a more expressive abstraction---say, \CTS{\Aff,\ints}---and use it for computing a composition in the closure algorithm,
widening to monotonicity constraints only at the level of cycles. This still allows the use of \MCTS algorithms for the bound analysis.
\eex

\section{Related Work}
\label{sec:rw}

There is a surprisingly large body of work related to the topics
of this paper. Most pertinent is the work in program analysis, directed at obtaining \emph{symbolic,
possibly asymptotic, complexity bounds} for programs (in a high-level language
or an intermediate language) under generic cost models (either unit cost or a more flexible, parametrized
cost model).
In this section, to put our work in context, we cite some of these works and indicate what approaches
were employed. The first subsection is an overview and cites various approaches. The second one
elaborates on the works most directly related to ours. There are many other works in this area which
have been left out; a complete survey would be an article in itself.

\subsection{Approaches in Complexity Analysis}

\paragraph{Seminal works.}
Wgbreit~\cite{Wegbreit:75} presented the first, and very influential, system for automatically analysing
a program's complexity. His system analyzes first-order LISP programs;
Broadly speaking, the system instruments a program to obtain a function that returns the desired
complexity measure, and then attempts to simplify the program until a closed form for the function
can be found. Possibly, the program becomes a set of recurrence equations for the complexity which
have to be solved.
Subsequent works along similar lines included \cite{ACE,Rosendahl89}
and more recently \cite{Benzinger01,Benzinger04} for functional programs and
\cite{DebrayLin93,Debray-sas94} for logic programs. The latter describe static analyses to deal with complications
particular to the semantics of logic programs, where programs compute sets of answers and 
involve backtracking.

\paragraph{Studies of restricted languages.}  Our approach in this paper involves the study of
complexity properties of a simplied, abstract program.  Research in \emph{Implicit Computational Complexity} (ICC) has produced
numerous examples of programming languages that are so restricted that they capture an intended
complexity class, that is, compute all, and only, functions of that class. Early examples include
\cite{Cobham:1964,KasaiAdachi:80,BC:92}. Many of these restrictions (e.g., \cite{Cobham:1964,BC:92}) may be seen (or are even
explicitly presented)  as imposing a certain type system on a language which, otherwise, could also
compute outside the intended complexity class; but this is not an \emph{automated analysis} in the
sense that the programmer has to supply the ``types" (in \cite{Cobham:1964}, and also some
later works like \cite{CraryWeirich:2000}, these are explicit resource bounds). In these cases one might
describe the technique more as \emph{certification} than analysis. However, ICC research has also
developed some methods that were later put to effective use in automated analysis. Two notable
examples are the method of term interpretations (see the paragraph on Term Rewriting Systems below),
and the method of \emph{linear types} \cite{Hofmann:iandc2003}, which yielded strong analysis techniques
as described, e.g., in \cite{Hoffmann:2010,Jost:2010}.

\paragraph{SPEED}
is an ambitious project from Microsoft Research to create a complexity analysis tool using
a variety of techniques, focusing on C programs~\cite{SPEED-CAV08,SPEED-POPL09,SPEED-PLDI09,SPEED-CAV09,Gulwani-PLDI10}.
In~\cite{SPEED-CAV08,SPEED-POPL09}, the essence of the technique is to instrument the program
with a counter, so that the desired resource usage becomes an output value, and bound this value
using invariant-generation methods.  In~\cite{SPEED-PLDI09}, the techniques are 
program transformation (called control-flow refinement) and ``progress invariants,"
which are used for obtaining more precise bounds for nested loops. In~\cite{Gulwani-PLDI10},
the term \emph{reachability bound} was coined.

\paragraph{Abstract interpretation techniques.} 
While abstract interpretation~\cite{Cousot:ACM:96} is the de-facto standard way of presenting many program analyses,
in the realm of complexity analysis its role has mostly been confined to supporting analyses (finding
the ranges of values etc).
As mentioned above, complexity analysis is sometimes reduced to computing a bound on computed
 values, and this is done by the traditional kind of abstract interpretation (invariant generation). 
However, there are a few works where abstract interpretations have been developed that
directly result in complexity properties.
In \cite{SAFE-FOPARA2010} it was done for space complexity of a functional language.
In \cite{NW06,JK08}, simple imperative programming languages have been analysed for complexity;
interestingly, because of the background in ICC rather than in static analysis, the terminology of
abstract interpretation is not used. These works were followed by
 \cite{BJK08,B2010:DICE} where it was shown that for languages of a similar style (imperative
 structure, very restricted in the usage of data, and non-deterministic in control flow except for 
 bounded loops), an abstract-interpretation based analysis is actually a \emph{decision procedure}:
 for example, one can \emph{decide} whether a program is polynomial-time. In this paper,
 we are also interested in abstract programs whose properties of interest are
 decidable. However, the nature of the abstract programs is very different.

\paragraph{Term Rewriting Systems}
are an elementary computational model that may be used to represent programs from
a variety of source languages.
There is already much work on complexity analysis for TRSs.
We mention two of the directions taken.
\cite{HirokawaMoser:cade08,HirokawaMoser:lpar08,AvanziniMoser:rta2009,Giesl-etal-cade2011}
employ the \emph{dependency pair method}, which like the model we are studying, was originally 
conceived for termination, and in fact has been effectively combined with size-change termination
\cite{TG:05,GTSF-JAR06,Codish_lazySCT}.

Another method that has extended its scope from proving
termination to proving complexity bounds in the context of Term Rewriting Systems
is the \emph{polynomial interpretation} method
\cite{BCMT:01},
later extended to other kinds of interpretation functions
\cite{MSW-fsttcs08,MarionPechoux09,BonfanteDeloup:2010,neurauter2010matrix,waldmann:2010}.
The method
has some resemblance to the analysis of transition systems with ranking functions, since the value
of an interpretation has to decrease as computation progresses, but interpretations have a particular
structure which is related to the structure of the terms in the system. Different interpretation methods
have very different structures and it is beyond the scope of this work to survey this line of work in greater
detail. It should be pointed out that, basically, interpretations are proof methods and it is not always clear
how to turn them into automatic analyses (in other words: how to \emph{synthesize} suitable 
interpretations), but this issue is discussed in the literature, for example in~\cite{Amadio05,MSW-fsttcs08} and many others.

\subsection{Analysis of Constraint Transition Systems.}

We have already described~\cite{ADFG:2010}, where  \CTS{\Aff,\ints} was used as an abstract
program and analysed using lexicographic linear ranking functions.

The COSTA project
~\cite{AAGP:jar2010,Albert-et-al:TCS:2011}  targets symbolic analysis of Java
bytecode programs.  It is a big project, in which involved methods of abstracting the concrete programs
were implemented, but this is unrelated to our topic. Our interest begins where they reach an
abstract program representation, which they call CRS (for \emph{cost relation system}). An example of 
a CRS (liberally modified from~\cite{AAGP:jar2010}) is:
\begin{align*}
&E(a,j) = k_1 + E(a',j') + F(a,j,j',a')  &&  \{ j'=j, a'=a-1, a'\ge 0, j\ge 0 \} \\
&F(a,j,j',a') = k_2 + E(a,j+1)  &&   \{  j < a-1,  j\ge 0, a-a'=1, j'=j \}
\end{align*}
where $k_1,k_2$ represents costs (and can be non-constant expressions depending on the variables);
essentially, this can be understood as a non-deterministic sort of recursive program whose result is the
desired cost bound. As a central part in the algorithm to bound this result, the system is simplified 
to eliminate indirect recursion (which is not possible for all systems, but is argued to work well in practice)
and then the height of the recursion tree is bounded by looking at individual (multiple-path) loops,
e.g., all the ``calls" from $E$ to $E$, and finding a linear ranking function for each such loop.
In a structured program with nested loops, each loop will turn into this kind of a recursive cost relation
and will therefore have to be bounded using a linear ranking function. This implies that a global ranking
function of the lexicographic linear kind exists, but the technique is more restricted than~\cite{ADFG:2010}
which finds a lexicographic linear ranking function by analysing the transition system
globally (that is, the lexicographic structure does not have to follow the loop nesting).

In comparison to our work, it is important to note that affine relations 
are expressive enough to make their termination problem undecidable (the simple argument
is that counter machines can be represented). Thus, a complete
solution cannot be achieved.
One could try to relate our works by considering \MCTS as a special case of \CTS{\Aff,\ints};
if we do so, we find that their solutions do not encompass ours as a special case. 
Indeed, not every \MCTS which is bounded terminating has a lexicographic linear ranking function
(not even systems with a single program point). This can be observed in some of our
examples, e.g., Example \ref{ex:nolex} in Section~\ref{sec:examples}.

\paragraph{Monotonicity constraint transition systems.} 
As mentioned earlier, monotonicity constraint transition systems have been first used (with different
terminology) for termination analysis of logic programs~\cite{Sa:91,LS:97,CT:99}. In addition to this
successful application, they have also been applied in the termination analysis of functional programs
\cite{leejonesbenamram01,MV-icse06,krauss07,SereniJones:05:hot} and imperative programs
\cite{SpotoMP09,Avery:06,CGBFG:11:TPLP}. Some works on the theory of \MCTS
and their decision problems are \cite{Codish-et-al:05,MT:09,Ben-Amram:ranking,BA:mcs,BA:mcInts} ;
decision procedures for extensions of the model have been discussed in \cite{BA:delta,BP2012}. 

While this paper was in preparation, two independnt works which also relate \MCTS
and cost analysis have been published.
 Zuleger et al.~\cite{ZGSV-sas11}
used \MCTS (more specifically, size-change graphs). The cited
conference paper does not provide all details, however even a superficial look confirms that their 
use of the abstraction is essentially different from our work, since they do not employ an ``abstract
and conquer'' approach where an abstract program becomes an object in itself. Instead, the abstraction
is just one tool in a complex algorithm that processes source programs, and is used to heuristically generate
ranking functions for loops. On the other hand, Bozzelli~\cite{Bozzelli:2012} analyses \emph{gap-constraint
transition systems}, in our notation,  \CTS{\Gap,\ints}. Gap constraints $\Gap$ are of the form $u \ge v+c$
where $u$, $v$ are (possibly primed) variables, $c\ge 0$ an integer, and it is also allowed to replace 
either $u$ or $v$ by a constant.  Clearly, this is an extension of \MCTS. The main result presented 
overlaps with ours; specifically, she proves the PSPACE complexity of bounded termination. 
An interesting result in~\cite{Bozzelli:2012} that we have not considered is how to compute a representation of the
initial states from which a system is bounded-terminating (when it is not always bounded). Bozzelli shows
that this can be represented using gap constraints. Clearly, this also holds for \MCTS.

\section{Conclusion}
\label{sec:conc}

The Monotonicity Constraint abstraction came into being specifically for the purpose
of termination analysis~\cite{CT:99,LS:97,Sa:91}.  It is natural to wish to
extend termination proofs
into complexity bounds. This work does it for the MC framework. For abstract programs, the 
complexity problem is to bound the length of transition sequences. Pleasantly, we find that the
problem is decidable, and its computational complexity is the same as termination. An interesting
conclusion is that a bound exists if and only if a polynomial one does (a different kind of statement than
stating that a certain analysis tool only finds polynomial bounds!). 

Since we are dealing with abstract programs, the question of relating these bounds to complexity of the
concrete program arises. We illustrate how the polynomial bound may mean polynomial time, space or
a polynomial exponent. In fact, classes PTIME, PSPACE and EXPTIME may all be captured by very
simple abstraction of programs to constraint systems.

We have not yet been able to perform an empirical evaluation, but at least theoretically,
our results sustain the claim that, just as they proved
quite useful for termination, MCs can contribute to complexity analysis.
To practically fulfill this promise, attention to the analysis of the concrete programs is necessary,
and to the scalability of the implementation.
 
In this work, we have not computed explicit bounds, and
we propose as an open problem the question whether
explicit bounds that are \emph{precise} can also be computed (in polynomial space?).

\bibliographystyle{plain}


\small

\begin{thebibliography}{10}

\bibitem{AAGP:sas08}
Elvira Albert, Puri Arenas, Samir Genaim, and Germ{\'a}n Puebla.
\newblock Automatic inference of upper bounds for recurrence relations in cost
  analysis.
\newblock In Mar{\'i}a Alpuente and Germ{\'a}n Vidal, editors, {\em Static
  Analysis, 15th International Symposium, {SAS} 2008, Valencia, Spain,
  Proceedings}, volume 5079 of {\em Lecture Notes in Computer Science}, pages
  221--237. Springer, 2008.

\bibitem{AAGP:jar2010}
Elvira Albert, Puri Arenas, Samir Genaim, and Germ\'an Puebla.
\newblock Closed-form upper bounds in static cost analysis.
\newblock {\em Journal of Automated Reasoning}, 46(2):161--203, 2010.

\bibitem{Albert-et-al:FSAD:2009}
Elvira Albert, Puri Arenas, Samir Genaim, Germ{\'a}n Puebla, and Damiano
  Zanardini.
\newblock Resource usage analysis and its application to resource
  certification.
\newblock In Alessandro Aldini, Gilles Barthe, and Roberto Gorrieri, editors,
  {\em Foundations of Security Analysis and Design V}, volume 5705 of {\em
  Lecture Notes in Computer Science}, pages 258--288. Springer Berlin /
  Heidelberg, 2009.

\bibitem{Albert-et-al:TCS:2011}
Elvira Albert, Puri Arenas, Samir Genaim, German Puebla, and Damiano Zanardini.
\newblock Cost analysis of object-oriented bytecode programs.
\newblock {\em Theoretical Computer Science}, 413(1):142--159, 2012.

\bibitem{ADFG:2010}
Christophe Alias, Alain Darte, Paul Feautrier, and Laure Gonnord.
\newblock Multi-dimensional rankings, program termination, and complexity
  bounds of flowchart programs.
\newblock In Radhia Cousot and Matthieu Martel, editors, {\em Static Analysis,
  Proceedings of the 17th International Symposium, Perpignan, France}, volume
  6337 of {\em Lecture Notes in Computer Science}, pages 117--133. Springer,
  2010.

\bibitem{Amadio05}
Roberto~M. Amadio.
\newblock Synthesis of max-plus quasi-interpretations.
\newblock {\em Fundam. Inform.}, 65(1-2):29--60, 2005.

\bibitem{AvanziniMoser:rta2009}
Martin Avanzini and Georg Moser.
\newblock Dependency pairs and polynomial path orders.
\newblock In Ralf Treinen, editor, {\em Rewriting Techniques and Applications,
  20th International Conference RTA 2009, Bras\'{\i}lia, Brazil, 2009,
  Proceedings}, volume 5595 of {\em LNCS 5595}, pages 48--62. Springer, 2009.

\bibitem{Avery:06}
James Avery.
\newblock Size-change termination and bound analysis.
\newblock In M.~Hagiya and P.~Wadler, editors, {\em Functional and Logic
  Programming: 8th International Symposium, FLOPS 2006}, volume 3945 of {\em
  Lecture Notes in Computer Science}. Springer, 2006.

\bibitem{BC:92}
Stephen Bellantoni and Stephen~A. Cook.
\newblock A new recursion-theoretic characterization of the polytime functions.
\newblock {\em Computational Complexity}, 2:97--110, 1992.

\bibitem{BA:delta}
Amir~M. Ben-Amram.
\newblock Size-change termination with difference constraints.
\newblock {\em ACM Trans. Program. Lang. Syst.}, 30(3):1--31, 2008.

\bibitem{Ben-Amram:ranking}
Amir~M. Ben-Amram.
\newblock A complexity tradeoff in ranking-function termination proofs.
\newblock {\em Acta Informatica}, 46(1):57--72, February 2009.

\bibitem{B2010:DICE}
Amir~M. Ben-Amram.
\newblock On decidable growth-rate properties of imperative programs.
\newblock In Patrick Baillot, editor, {\em International Workshop on
  Developments in Implicit Computational complExity (DICE 2010)}, volume~23 of
  {\em EPTCS}, pages 1--14, 2010.

\bibitem{BA:mcs}
Amir~M. Ben-Amram.
\newblock Size-change termination, monotonicity constraints and ranking
  functions.
\newblock {\em Logical Methods in Computer Science}, 6(3), 2010.

\bibitem{BA:mcInts}
Amir~M. Ben-Amram.
\newblock Monotonicity constraints for termination in the integer domain.
\newblock {\em Logical Methods in Computer Science}, 7:1--43, 2011.

\bibitem{BC:08:TACAS}
Amir~M. Ben-Amram and Michael Codish.
\newblock A {SAT}-based approach to size change termination with global ranking
  functions.
\newblock In C.R. Ramakrishnan and Jakob Rehof, editors, {\em 14th Intl.
  Conference on Tools and Algorithms for the Construction and Analysis of
  Systems ({TACAS})}, volume 5028 of {\em LNCS}, pages 46--55. Springer, 2008.

\bibitem{BJK08}
Amir~M. {Ben-Amram}, Neil~D. Jones, and Lars Kristiansen.
\newblock Linear, polynomial or exponential? complexity inference in polynomial
  time.
\newblock In Arnold Beckmann, Costas Dimitracopoulos, and Benedikt L{\"o}we,
  editors, {\em Logic and Theory of Algorithms, Fourth Conference on
  Computability in Europe, {CiE} 2008}, volume 5028 of {\em LNCS}, pages
  67--76. Springer, 2008.

\bibitem{Benzinger01}
Ralph Benzinger.
\newblock Automated complexity analysis of nuprl extracted programs.
\newblock {\em Journal of Functional Programming}, 11(1):3--31, 2001.

\bibitem{Benzinger04}
Ralph Benzinger.
\newblock Automated higher-order complexity analysis.
\newblock {\em Theoretical Computer Science}, 318(1-2):79--103, 2004.

\bibitem{BCMT:01}
Guillaume Bonfante, Adam Cichon, Jean-Yves Marion, and H{\'e}l{\`e}ne Touzet.
\newblock Algorithms with polynomial interpretation termination proof.
\newblock {\em Journal of Functional Programming}, 11:33--53, 2001.

\bibitem{BonfanteDeloup:2010}
Guillaume Bonfante and Florian Deloup.
\newblock Complexity invariance of real interpretations.
\newblock In Jan Kratochv{\'\i}l, Angsheng Li, Jir{\'\i} Fiala, and Petr
  Kolman, editors, {\em Theory and Applications of Models of Computation},
  volume 6108 of {\em Lecture Notes in Computer Science}, pages 139--150.
  Springer Berlin / Heidelberg, 2010.

\bibitem{Bozzelli:2012}
Laura Bozzelli.
\newblock Strong termination for gap-order constraint abstractions of counter
  systems.
\newblock In Adrian~Horia Dediu and Carlos Mart{\'i}n-Vide, editors, {\em
  Language and Automata Theory and Applications - 6th International Conference,
  ({LATA})}, volume 7183 of {\em Lecture Notes in Computer Science}, pages
  155--168. Springer, 2012.

\bibitem{BP2012}
Laura Bozzelli and Sophie Pinchinat.
\newblock Verification of gap-order constraint abstractions of counter systems.
\newblock {\em Theoretical Computer Science}, 523(0):1 -- 36, 2014.

\bibitem{CichonLescanne:1992}
Adam Cichon and Pierre Lescanne.
\newblock Polynomial interpretations and the complexity of algorithms.
\newblock In {\em Proceedings of the 11th International Conference on Automated
  Deduction: Automated Deduction}, CADE-11, pages 139--147. Springer-Verlag,
  1992.

\bibitem{Cobham:1964}
Alan~B. Cobham.
\newblock The intrinsic computational difficulty of functions.
\newblock In Y.~Bar-Hillel, editor, {\em Proceeding of the~1964 International
  Congress for Logic, Methodology, and Philosophy of Science}, pages 24--30.
  North-Holland, Amsterdam, 1964.

\bibitem{Codish_lazySCT}
Michael Codish, Carsten Fuhs, J{\"u}rgen Giesl, and Peter Schneider-kamp.
\newblock Lazy abstraction for size-change termination.
\newblock In Christian~G. Ferm{\"u}ller and Andrei Voronkov, editors, {\em 17th
  International Conference on Logic for Programming, Artificial Intelligence
  and Reasoning (LPAR 2010)}, volume 6397 of {\em Lecture Notes in Computer
  Science}, pages 217--232. Springer, 2010.

\bibitem{CGBFG:11:TPLP}
Michael Codish, Igor Gonopolskiy, Amir~M. Ben-Amram, Carsten Fuhs, and
  J{\"u}rgen Giesl.
\newblock {SAT}-based termination analysis using monotonicity constraints over
  the integers.
\newblock {\em Theory and Practice of Logic Programming, 26th Int'l. Conference
  on Logic Programming (ICLP'11) Special Issue}, 11(Special Issue
  4-5):503--520, July 2011.

\bibitem{Codish-et-al:05}
Michael Codish, Vitaly Lagoon, and Peter~J. Stuckey.
\newblock Testing for termination with monotonicity constraints.
\newblock In Maurizio Gabbrielli and Gopal Gupta, editors, {\em Logic
  Programming, 21st International Conference, ICLP 2005}, volume 3668 of {\em
  Lecture Notes in Computer Science}, pages 326--340. Springer, 2005.

\bibitem{CT:99}
Michael Codish and Cohavit Taboch.
\newblock A semantic basis for termination analysis of logic programs.
\newblock {\em The Journal of Logic Programming}, 41(1):103--123, 1999.
\newblock preliminary (conference) version in LNCS 1298 (1997).

\bibitem{ADFG:web}
{LIP} Compsys~Team.
\newblock {WTC} toolsuite benchmarks.
\newblock \url{http://www.ens-lyon.fr/LIP/COMPSYS/Tools/Ranking/}, September
  2010.

\bibitem{Cousot:ACM:96}
Patrick Cousot.
\newblock Abstract interpretation.
\newblock {\em ACM Computing Surveys}, 28(2):324--328, 1996.

\bibitem{CousotHalbwachs:1978}
Patrick Cousot and Nicholas Halbwachs.
\newblock Automatic discovery of linear restraints among variables of a
  program.
\newblock In {\em Conference Record of the Fifth annual {ACM} Symposium on
  Principles of Programming Languages}, pages 84--96. ACM, ACM, January 1978.

\bibitem{CraryWeirich:2000}
Karl Crary and Stephnie Weirich.
\newblock Resource bound certification.
\newblock In {\em Proceedings of the 27th ACM SIGPLAN-SIGACT symposium on
  Principles of programming languages}, POPL '00, pages 184--198, New York, NY,
  USA, 2000. ACM.

\bibitem{Debray-sas94}
Saumya Debray, Pedro L\'opez-Garc\'ia, Manuel~V. Hermenegildo, and Nai wei Lin.
\newblock Estimating the computational cost of logic programs.
\newblock In {\em First International Static Analysis Symposium, SAS'94 Namur,
  Belgium, Proceedings}, pages 255--265. LNCS 864, 1994.

\bibitem{DebrayLin93}
Saumya Debray and Nai wei Lin.
\newblock Cost analysis of logic programs.
\newblock {\em ACM Transactions on Programming Languages and Systems},
  15:48--62, 1993.

\bibitem{GTSF-JAR06}
J.~Giesl, R.~Thiemann, P.~Schneider-Kamp, and S.~Falke.
\newblock Mechanizing and improving dependency pairs.
\newblock {\em Journal of Automated Reasoning}, 37(3):155--203, 2006.

\bibitem{SPEED-CAV08}
Bhargav~S. Gulavani and Sumit Gulwani.
\newblock A numerical abstract domain based on expression abstraction and max
  operator with application in timing analysis.
\newblock In {\em Computer Aided Verification, 20th International Conference,
  CAV 2008, Princeton, NJ, USA, July 7-14, 2008, Proceedings}, volume 5123 of
  {\em Lecture Notes in Computer Science}, pages 370--384. Springer, 2008.

\bibitem{SPEED-CAV09}
Sumit Gulwani.
\newblock {SPEED}: Symbolic complexity bound analysis.
\newblock In Ahmed Bouajjani and Oded Maler, editors, {\em Computer Aided
  Verification, 21st International Conference, {CAV} 2009, Grenoble, France},
  volume 5643 of {\em Lecture Notes in Computer Science}, pages 51--62.
  Springer, 2009.

\bibitem{SPEED-PLDI09}
Sumit Gulwani, Sagar Jain, and Eric Koskinen.
\newblock Control-flow refinement and progress invariants for bound analysis.
\newblock In Michael Hind and Amer Diwan, editors, {\em Proceedings of the 2009
  {ACM} {SIGPLAN} Conference on Programming Language Design and Implementation,
  {PLDI} 2009, Dublin, Ireland, June 15-21, 2009}, pages 375--385. ACM, 2009.

\bibitem{SPEED-POPL09}
Sumit Gulwani, Krishna~K. Mehra, and Trishul~M. Chilimbi.
\newblock {SPEED}: precise and efficient static estimation of program
  computational complexity.
\newblock In Zhong Shao and Benjamin~C. Pierce, editors, {\em Proceedings of
  the 36th {ACM} {SIGPLAN}-{SIGACT} Symposium on Principles of Programming
  Languages, {POPL} 2009, Savannah, {GA}, {USA}}, pages 127--139. ACM, 2009.

\bibitem{Gulwani-PLDI10}
Sumit Gulwani and Florian Zuleger.
\newblock The reachability-bound problem.
\newblock In Benjamin~G. Zorn and Alexander Aiken, editors, {\em Proceedings of
  the 2010 {ACM} {SIGPLAN} Conference on Programming Language Design and
  Implementation, {PLDI} 2010, Toronto, Ontario, Canada, June 5-10, 2010},
  pages 292--304. ACM, 2010.

\bibitem{HirokawaMoser:lpar08}
N.~Hirokawa and G.~Moser.
\newblock Complexity, graphs, and the dependency pair method.
\newblock In {\em Proc.\ LPAR~'08}, volume 5330 of {\em LNAI}, pages 652--666,
  2008.

\bibitem{HirokawaMoser:cade08}
Nao Hirokawa and Georg Moser.
\newblock Automated complexity analysis based on the dependency pair method.
\newblock In Alessandro Armando, Peter Baumgartner, and Gilles Dowek, editors,
  {\em Automated Reasoning, 4th International Joint Conference, {IJCAR} 2008,
  Sydney, Australia, August 12-15, 2008, Proceedings}, volume 5195 of {\em
  Lecture Notes in Computer Science}, pages 364--379. Springer, 2008.
\newblock Full version available on-line (under review).

\bibitem{Hoffmann:2010}
Jan Hoffmann and Martin Hofmann.
\newblock Amortized resource analysis with polynomial potential.
\newblock In Andrew Gordon, editor, {\em Programming Languages and Systems},
  volume 6012 of {\em Lecture Notes in Computer Science}, pages 287--306.
  Springer Berlin / Heidelberg, 2010.

\bibitem{Hofmann:iandc2003}
Martin Hofmann.
\newblock Linear types and non-size-increasing polynomial time computation.
\newblock {\em Information and Computation}, 183(1):57--85, 2003.
\newblock Special Issue: International Workshop on Implicit Computational
  Complexity (ICC'99).

\bibitem{JK08}
Neil~D. Jones and Lars Kristiansen.
\newblock A flow calculus of mwp-bounds for complexity analysis.
\newblock {\em ACM Trans. Computational Logic}, 10(4):1--41, 2009.

\bibitem{Jost:2010}
Steffen Jost, Kevin Hammond, Hans-Wolfgang Loidl, and Martin Hofmann.
\newblock Static determination of quantitative resource usage for higher-order
  programs.
\newblock In {\em The 37th annual ACM SIGPLAN-SIGACT symposium on Principles of
  Programming Languages}, (POPL), pages 223--236, New York, NY, USA, 2010. ACM.

\bibitem{KasaiAdachi:80}
Takumi Kasai and Akeo Adachi.
\newblock A characterization of time complexity by simple loop programs.
\newblock {\em Journal of Computer and System Sciences}, 20(1):1--17, 1980.

\bibitem{krauss07}
Alexander Krauss.
\newblock Certified size-change termination.
\newblock In Frank Pfenning, editor, {\em 11th International Conference on
  Automated Deduction (CADE)}, volume 4603 of {\em LNAI}, pages 460--475.
  Springer-Verlag, July 2007.

\bibitem{KraussHeller09}
Alexander Krauss and Armin Heller.
\newblock A mechanized proof reconstruction for {SCNP} termination.
\newblock Presented in the Tenth International Workshop on Termination WST'09,
  Leipzig, 2009.

\bibitem{ACE}
Daniel Le~M\'{e}tayer.
\newblock Ace: an automatic complexity evaluator.
\newblock {\em ACM Trans. Program. Lang. Syst.}, 10(2):248--266, 1988.

\bibitem{leejonesbenamram01}
Chin~Soon Lee, Neil~D. Jones, and Amir~M. Ben-Amram.
\newblock The size-change principle for program termination.
\newblock In {\em Proceedings of the Twenty-Eigth {ACM} Symposium on Principles
  of Programming Languages, January 2001}, volume~28, pages 81--92. ACM press,
  January 2001.

\bibitem{LS:97}
Naomi Lindenstrauss and Yehoshua Sagiv.
\newblock Automatic termination analysis of {P}rolog programs.
\newblock In Lee Naish, editor, {\em Proceedings of the Fourteenth
  International Conference on Logic Programming}, pages 64--77, {L}euven,
  {B}elgium, Jul 1997. {MIT} {P}ress.

\bibitem{MT:09}
Panagiotis Manolios and Aaron Turon.
\newblock {All-Termination(T)}.
\newblock In {\em Proceedings of the 15th Intl. Conference on Tools and
  Algorithms for the Construction and Analysis of Systems ({TACAS})}, volume
  5505 of {\em Lecture Notes in Computer Science}, pages 398--412. Springer,
  2009.

\bibitem{MV-icse06}
Panagiotis Manolios and Daron Vroon.
\newblock Integrating static analysis and general-purpose theorem proving for
  termination analysis.
\newblock In {\em Proceedings of ICSE'06: International Conference on Software
  Engineering}, pages 873--876. ACM, 2006.

\bibitem{MV-cav06}
Panagiotis Manolios and Daron Vroon.
\newblock Termination analysis with calling context graphs.
\newblock In {\em Proceedings, Computer Aided Verification, 18th International
  Conference, {CAV} 2006, {S}eattle, {WA}, {USA}}, volume 4144 of {\em LNCS},
  pages 401--414. Springer-Verlag, 2006.

\bibitem{MarionPechoux09}
Jean-Yves Marion and Romain P{\'e}choux.
\newblock Sup-interpretations, a semantic method for static analysis of program
  resources.
\newblock {\em {ACM} Transactions on Computational Logic}, 10(4), 2009.

\bibitem{MR:67}
Albert~R. Meyer and Dennis~M. Ritchie.
\newblock The complexity of loop programs.
\newblock In {\em Proc. 22nd ACM National Conference}, pages 465--469,
  Washington, DC, 1967.

\bibitem{SAFE-FOPARA2010}
Manuel Montenegro, Ricardo Pe\~na, and Clara Segura.
\newblock A space consumption analysis by abstract interpretation.
\newblock In Marko van Eekelen and Olha Shkaravska, editors, {\em Foundational
  and Practical Aspects of Resource Analysis}, volume 6324 of {\em Lecture
  Notes in Computer Science}, pages 34--50. Springer Berlin / Heidelberg, 2010.

\bibitem{MSW-fsttcs08}
G.~Moser, A.~Schnabl, and J.~Waldmann.
\newblock Complexity analysis of term rewriting based on matrix and context
  dependent interpretations.
\newblock In Ramesh Hariharan, Madhavan Mukund, and V~Vinay, editors, {\em
  Proc.\ FSTTCS~'08}, LIPIcs 2, pages 304--315. Dagstuhl Publishing, 2008.

\bibitem{neurauter2010matrix}
F.~Neurauter, H.~Zankl, and A.~Middeldorp.
\newblock Revisiting matrix interpretations for polynomial derivational
  complexity of term rewriting.
\newblock In {\em Proc.\ LPAR~'10}, volume 6397 of {\em LNCS}, pages 550--564,
  2010.

\bibitem{NW06}
Karl-Heinz Niggl and Henning Wunderlich.
\newblock Certifying polynomial time and linear/polynomial space for imperative
  programs.
\newblock {\em SIAM J. Comput}, 35(5):1122--1147, 2006.

\bibitem{Giesl-etal-cade2011}
Lars Noschinski, Fabian Emmes, and J{\"u}rgen Giesl.
\newblock A dependency pair framework for innermost complexity analysis of term
  rewrite systems.
\newblock In Nikolaj Bj{\o}rner and Viorica Sofronie-Stokkermans, editors, {\em
  Automated Deduction, 23rd International Conference, Wroclaw, Poland, July 31
  - August 5, 2011. Proceedings}, volume 6803 of {\em Lecture Notes in Computer
  Science}, pages 422--438. Springer, 2011.

\bibitem{Rosendahl89}
M.~Rosendahl.
\newblock Automatic complexity analysis.
\newblock In {\em Proceedings of the Conference on Functional Programming
  Languages and Computer Architecture {FPCA}'89}, pages 144--156. ACM, 1989.

\bibitem{Sa:91}
Yehoshua Sagiv.
\newblock A termination test for logic programs.
\newblock In Vijay Saraswat and Kazunori Ueda, editors, {\em Logic Programming,
  Proceedings of the 1991 International Symposium, {S}an {D}iego, {C}alifornia,
  {USA}}, pages 518--532. {MIT} {P}ress, 1991.

\bibitem{SereniJones:05:hot}
Damien Sereni and Neil~D. Jones.
\newblock Termination analysis of higher-order functional programs.
\newblock In {\em Proceedings of the Third Asian Symposium on Programming
  Languages and Systems (APLAS 2005)}, volume 3780 of {\em Lecture Notes in
  Computer Science}, pages 281--297. Springer-Verlag, 2005.

\bibitem{SpotoMP09}
Fausto Spoto, Fred Mesnard, and \'{E}tienne Payet.
\newblock A termination analyzer for {Java} bytecode based on path-length.
\newblock {\em ACM Trans. Program. Lang. Syst.}, 32(3):1--70, 2010.

\bibitem{TG:05}
Ren{\'e} Thiemann and J{\"u}rgen Giesl.
\newblock The size-change principle and dependency pairs for termination of
  term rewriting.
\newblock {\em Applicable Algebra in Engineering, Communication and Computing},
  16(4):229--270, September 2005.

\bibitem{waldmann:2010}
Johannes Waldmann.
\newblock Polynomially bounded matrix interpretations.
\newblock In Christopher Lynch, editor, {\em Proceedings of the 21st
  International Conference on Rewriting Techniques and Applications}, volume~6
  of {\em Leibniz International Proceedings in Informatics (LIPIcs)}, pages
  357--372, Dagstuhl, Germany, 2010. Schloss Dagstuhl--Leibniz-Zentrum fuer
  Informatik.

\bibitem{Wegbreit:75}
Ben Wegbreit.
\newblock Mechanical program analysis.
\newblock {\em Communications of the ACM}, 18(9):528--539, 1975.

\bibitem{ZGSV-sas11}
Florian Zuleger, Sumit Gulwani, Moritz Sinn, and Helmut Veith.
\newblock Bound analysis of imperative programs with the size-change
  abstraction.
\newblock In Eran Yahav, editor, {\em Proceedings of the 8th International
  Symposium on Static Analysis, SAS 2011, Venice, Italy}, volume 6887 of {\em
  Lecture Notes in Computer Science}, pages 280--297, 2011.

\end{thebibliography}

\end{document}